%% file: quop_mpi.tex
\documentclass[preprint,5p,times,twocolumn]{elsarticle}
\usepackage{stmaryrd}
\usepackage{wrapfig}
\usepackage[export]{adjustbox}
\usepackage{amsmath}
\usepackage{amssymb}
\usepackage{bm}
\usepackage{dsfont}
\usepackage{array, makecell}
\usepackage{graphicx, adjustbox}
\usepackage{mathtools}
\usepackage{setspace}
\usepackage{lipsum}  
\usepackage{cprotect}
\usepackage{textpos}

\usepackage{tikz}
\usetikzlibrary{positioning,shapes.geometric}
\usetikzlibrary{shapes, positioning, calc, arrows.meta,decorations.pathreplacing,calligraphy}

\usepackage{xurl} 
\usepackage{hyperref}
\usepackage{lineno}
\usepackage{enumitem}
\usepackage{amsfonts}
\usepackage{placeins}
\usepackage[capitalise,nameinlink,noabbrev]{cleveref}
\usepackage{textcomp}
\usepackage{subcaption}
\usepackage{soul}
\usepackage{multicol}
\usepackage{float}
\usepackage{wrapfig}

\usepackage{mathtools}

\newcommand{\colvec}[1]{\begin{pmatrix}   #1 \end{pmatrix}}
 
\usepackage[most]{tcolorbox}

\newtcbox{\class}[1][]{enhanced,
 box align=base,
 nobeforeafter,
 colback=white,
 colframe={lightgray},
 size=small,
 left=0pt,
 right=0pt,
 boxsep=1.5pt,
 #1}
 
 \newtcbox{\submodule}[1][]{enhanced,
 box align=base,
 nobeforeafter,
 colback=white,
 colframe={lightgray},
 size=small,
 left=0pt,
 right=0pt,
 boxsep=1.5pt,
 sharp corners,
 borderline={0.1pt}{1.8pt}{lightgray},
 #1}
 
 \newtcbox{\subsubmodule}[1][]{enhanced,
 box align=base,
 nobeforeafter,
 colback=white,
 colframe={lightgray},
 size=small,
 left=0pt,
 right=0pt,
 boxsep=1.5pt,
 borderline={0.1pt}{1.8pt}{lightgray},
 #1}

\usepackage{xcolor}
\PassOptionsToPackage{svgnames,x11names,dvipsnames}{xcolor}

\input{definitions.tex}

\renewenvironment{quote}
  {\small\list{}{\rightmargin=0cm \leftmargin=0.5cm}%
   \item\relax}
  {\endlist}

\makeatletter
\newcommand{\customlabel}[2]{%
   \protected@write \@auxout {}{\string \newlabel {#1}{{#2}{\thepage}{#2}{#1}{}} }%
   \hypertarget{#1}{}%
}%

\makeatother

\biboptions{sort&compress}

\begin{document}

\begin{frontmatter}

\title{\quop{}: a framework for parallel simulation of quantum variational algorithms.}

\author[a]{Edric Matwiejew\corref{author}}
\author[a]{Jingbo B. Wang}

\cortext[author]{Corresponding author.\\\textit{E-mail address:} Edric.Matwiejew@research.uwa.edu.au}
\address[a]{Department of Physics, The University of Western Australia, Perth, Australia}

\begin{abstract} 
\quop{} is a \python{} package designed for parallel simulation of quantum variational algorithms. It presents an object-orientated approach to quantum variational algorithm design and utilises MPI-parallelised sparse-matrix exponentiation, the fast Fourier transform and parallel gradient evaluation to achieve the highly efficient simulation of the fundamental unitary dynamics on massively parallel systems. In this article, we introduce \quop{} and explore its application to the simulation of quantum algorithms designed to solve combinatorial optimisation algorithms, including the Quantum Approximation Optimisation Algorithm, the Quantum Alternating Operator Ansatz, and the Quantum Walk-assisted Optimisation Algorithm.
\end{abstract}


\end{frontmatter}

\raggedbottom
\section{Introduction}

With the first generation of quantum computers currently in operation, the start of a new computing paradigm appears just around the corner \cite{Matthews_2021, Cerezo_2021}. Contributing to this optimism has been the development of algorithms that exploit a combination of classical and quantum hardware to solve optimisation problems \cite{farhi_quantum_2014, hadfield_quantum_2019,marsh_combinatorial_2019, marsh_combinatorial_2020,guerreschi_practical_2017}. Compared to many exclusively quantum algorithms, these quantum variational algorithms (QVAs) require minimal quantum operations and are inherently resilient to system noise \cite{peruzzo_variational_2014, Cerezo_2021}. For these reasons, QVAs are strong contenders for early practical applications of quantum computing in the Noisy Intermediate Scale Quantum (NISQ) era \cite{preskill_quantum_2018}. Examples of such QVAs include the Quantum Approximate Optimisation Algorithm  (\qaoa{}) \cite{farhi_quantum_2014,guerreschi_practical_2017}, the Quantum Alternating Operator Ansatz (\qaoaz{}) \cite{hadfield_quantum_2019}, and the Quantum Walk-assisted Optimisation Algorithm (\qwoa{}) \cite{marsh_combinatorial_2019, marsh_combinatorial_2020}, which have been designed to find optimal, or near-optimal, solutions to combinatorial optimisation problems. 

Combinatorial optimisation problems (COPs) —the task of finding the best combination of items from a set—are nearly ubiquitous \cite{kell_scientific_2012}. They are present in fields such as logistics \cite{sanchez_comparative_2020}, drug design \cite{liu_combinatorial_2017}, software compilation \cite{lozano_combinatorial_2019} and finance \cite{markowitz_portfolio_1952, palczewski_lp_2018}. These problems are difficult to solve classically due to a lack of identifiable structure and exponential growth of the solution space. Quantum variational algorithms can provide a polynomial speedup compared to a classical random search \cite{marsh_combinatorial_2019, marsh_combinatorial_2020} which is an attractive prospect in problems with great humanitarian or financial consequence. 

To solve COPs, QVAs exploit quantum superposition to act on the entire problem-specific solution space in quantum parallel. They apply a sequence of alternating unitaries; the first encodes the solution `qualities' into the phase of superposed quantum states, and the second `mixes' probability amplitude between the states. The phase-shift and mixing unitaries are parameterised by scalar variables adjusted iteratively by a classical optimiser that minimises the average measured solution quality. By encoding optimal solutions as minima in the solution space, lowering the average solution quality corresponds to amplifying probability density at quantum states associated with optimal or near-optimal solutions.  

Classical numerical simulation plays a vital role in the development of QVAs. Through simulation of the idealised quantum dynamics, researchers can study QVAs independently of implementation-specific hardware constraints and at scales that still exceed the functional limitations of current quantum hardware \cite{willsch_gpu-accelerated_2021}. To assist with these efforts, we have developed \quop{} (\textbf{Qu}antum \textbf{Op}timisation with \textbf{MPI}) \cite{edric_quop_mpi}, which provides a flexible framework for the design and classical simulation of QVAs. 

There is currently significant interest in developing tools for the simulations of QVAs in a high-performance computing setting. Recent examples include TensorFlow Quantum, a software framework for quantum machine learning \cite{broughton_tensorflow_2020}, and the J\"{u}lich universal quantum computer simulator \cite{willsch_gpu-accelerated_2021}. Both utilise GPU acceleration, with the latter being targeted at distributed GPU clusters. Also of note is the XACC framework and qFlex, which utilise a tensor network approach to quantum simulation \cite{mccaskey_xacc_2020,villalonga_flexible_2019}. These packages take a quantum-gate based approach to algorithm simulation and can simulate QVAs with a large number of qubits (e.g. more than 50) given a quantum circuit structure that is parsimonious to their underlying simulation methods. \quop{} presents a distinct option for QVA simulation in that it does not take a gate-based or approximative approach; instead, it focuses on the simulation of the fundamental unitary dynamics across the complete quantum state-space. It also provides the first tool for ready-made simulation of the Quantum Walk-assisted Optimisation Algorithm.

The structure of the paper is as follows. In \cref{sec:theoretical} we define the generalised QVA, introduce the \qaoa{}, \qaoaz{} and \qwoa{}, and specify the problem of combinatorial optimisation. In \cref{sec:numerical_methods,sec:parallel} we discuss the data structures, algorithms and parallelisation schemes leveraged by \quop{}. This is followed by an overview of the package structure, functionality and workflow types in \cref{sec:overview}. In \cref{sec:usage} we provide usage examples drawn from literature followed by a discussion of the package's performance in \cref{sec:performence}. Finally, concluding statements are presented in \cref{sec:conclusion}.

\section{Theoretical Background} \label{sec:theoretical}

\subsection{Quantum Variational Algorithms}\label{sec:QVA}

For a quantum system of size $N = 2^n$, where integer $n$ is a number of qubits with basis states $\set{\ket{0} = \colvec{0 \\ 1}, \ket{1} = \colvec{1 \\ 0}}$, \quop{} defines a generalised QVA as

\begin{equation}
    \label{eq:variational_generic}
    \ansatz{}=\left( \prod_{i = 1}^{D}\U{}(\theta_i) \right)\initialState{},
\end{equation}

\noindent where $\initialState{} \in \mathbb{C}^{N}$ is an initial quantum state with basis states $\set{\ket{i}}_{i=0,...,N-1}$, $\U{} \in \mathbb{C}^{N \times N}$ is the ansatz\footnote{Originating from a German word that refers to the starting thought of a process. In mathematics, an ansatz is an educated guess or assumption made to help solve a problem.} unitary , integer $D \geq 0$ specifies the number of applications of $\U{}$ to $\initialState{}$ (the `depth') and $\variationalParameters = \{ \theta_i \in \mathbb{R} \}$ is an ordered set of classically tunable values that parameterise $\U{}$. The ansatz unitary $\U{}$ and initial quantum state $\initialState{}$ together define a specific QVA.

A Quantum Variational Algorithm is executed by repeatedly preparing $\ansatz{}$ and measuring the expectation value
  
\begin{equation}
    \label{eq:objective_generic}
    \objectiveFunction{}(\variationalParameters) = \bra{\variationalParameters} \qualityOperator \ket{\variationalParameters}_\text{},
\end{equation}

\noindent where $\qualityOperator \in \mathbb{R}^{N \times N}$ is a diagonal matrix operator with entries $\qualityVector{} = q_i$ that specify the `quality' associated with quantum state $\ket{i}$. The variational parameters $\variationalParameters{}$ are updated using a classical optimiser with the objective being minimisation of $\objectiveFunction{}$.

The ansatz operator $\U{}$ specifies a sequence of alternating unitaries. This can include phase-shifts 

\begin{equation}
    \label{eq:phase_shift_generic}
    \U{phase}(\gamma) = \exp(-\text{i} \gamma \hat{O}),
\end{equation}

\noindent where $\hat{O} = \sum_{i=0}^{N-1} o_i \ket{i}\bra{i}$ is a diagonal phase-shift matrix operator, $\gamma \in \variationalParameters{}$ and $\hat{U}_\text{phase}$ applies a phase-shift proportional to $o_i$, as well as mixing-unitaries

\begin{equation}
    \label{eq:mixing_generic}
    \U{mix}(t) = \exp(-\text{i} t \hat{W}),
\end{equation}

\noindent where $t \in \variationalParameters{}$ is non-negative and $\hat{W} = \sum_{{i,j} = 0}^{n-1}w_{ij}\ket{j}\bra{i}$ is a mixing matrix operator in which non-diagonal entries specify coupling between states $\ket{i}$ and $\ket{j}$. Mixing-unitaries $\hat{U}_\text{mix}$ drive the transfer of probability amplitude between quantum states, during which encoded phase differences contribute to constructive and destructive interference. 

Phase-shift operators $\hat{O}$ and mixing operators $\hat{W}$ may also be parameterised by $\variationalParameters{}$. As these operators are time-independent Hamiltonians of the time-evolution operator, changes to the corresponding $\theta_i$ alter the element-wise magnitudes or structure of the matrix exponent before preparation of $\ansatz{}$. 

Typically, the ansatz unitary $\U{}$ is applied $D$ times to $\initialState{}$ with each repetition parameterised by subset $\theta \subseteq \variationalParameters{}$. Doing so increases the potential for constructive and destructive inference to concentrate probability amplitude at high-quality solutions; at the expense of classical optimisation over a larger parameter space and a deeper quantum circuit. In practice, a QVA must balance the improved convergence afforded by increases to $D$ against the ability of the quantum hardware to maintain coherence over a longer sequence of quantum operations.

Sections \ref{sec:unconstrainedOptimisation} and \ref{sec:constrainedOptimisation} introduce four distinct QVAs for solving constrained and unconstrained COPs. We summarise here the following notational conventions for a given QVA:

\begin{itemize}
	\item $n$: the number of qubits.
	\item $\initialState{}$: the initial quantum state vector.
	\item $\U{}$: a sequence of phase-shift and mixing operators.
	\item $\ket{\psi}$: $\initialState{}$ after $D \geq 0$ applications of $\U{}$.
	\item $\ansatz{}$: $\initialState{}$ after $D \geq 1$ applications of $\U{}$.
	\item $\variationalParameters{}$: classically tunable variables parameterising $\U{}$ with starting values $\variationalParameters_0$ and optimised values $\variationalParameters_f$.
	\item $\objectiveFunction{}$: the ansatz objective function.
\end{itemize} 

\subsection{Combinatorial Optimisation with QVAs} \label{sec:CO}

Combinatorial optimisation problems seek optimal solutions $\optimalSolution{}$ of the form,

\begin{equation}
	\optimalSolution = \Big\{\solution \; | \; \costFunction{\solution} \in  \min \set{ \costFunction{\solution} \; | \; \solution \in \validSolutionSpace{}}\Big\},
\end{equation}

\noindent where the problem cost-function $\costFunction{\solution{}}$ maps a solution $\solution{}$ from an ordered set of problem solutions $\mathcal{S} = \set{\solution_i}$ to $\mathbb{R}$, $s$ is a $k$-permutation of discrete elements from a finite set $\solutionElements{}$ and

\begin{equation}
	\validSolutionSpace = \set{ \solution \; | \; \solution \in \solutionConstraints}
\end{equation} 

\noindent is the problem-specific valid solution space where

\begin{equation}
 \solutionConstraints = \bigcup_i \Big\{\solution \; | \; \chi_i(\solution) = a_i\Big\}   
\end{equation}

\noindent denotes any constraints on $\optimalSolution{}$ and $\bm{a} = \set{a_i}$ defines the constraints. 

Problems of this type are often difficult to solve as $\solutionSpace{}$ grows factorially with $\size{\solutionElements{}}$ and, in general, lacks identifiable structure. For this reason, heuristic and metaheuristic algorithms are often used to find solutions that satisfy the relaxed condition of $\costFunction{\optimalSolution{}}$ being a `sufficiently low' local minimum.

To apply a quantum variational algorithm to a given combinatorial optimisation problem, an injective map is defined between $\solutionSpace{}$ and $\mathcal{H}$  with the cost-function values forming the diagonal of the quality operator $\qualityVector = \costFunction{\solution_i}$. For example, a problem with four solutions, $\mathcal{S} = \{s_0, s_1, s_2, s_3\}$, maps to a two-qubit system as

\begin{equation}
	\begin{aligned}
		\ket{00} &= \ket{0} \rightarrow \ket{s_0} \\
		\ket{01} &= \ket{1} \rightarrow \ket{s_1} \\
		\ket{10} &= \ket{2} \rightarrow \ket{s_2} \\
		\ket{11} &= \ket{3} \rightarrow \ket{s_3},
	\end{aligned}
\end{equation}   

\noindent where $\qualityVector = \Big(C(s_0),C(s_1),C(s_2),C(s_3)\Big)$.

For a combinatorial optimisation problem to be efficiently solvable by a QVA, it must satisfy three conditions:

\begin{enumerate}
	\item The number of solutions $\size{\solutionSpace}$ must be efficiently computable in order to establish a bound on the size of the required Hilbert space $\mathcal{H}$. 
	\item For all solutions $s$, $\costFunction{\solution}$ must be computable in polynomial time .
	\item For all solutions $s$, $\costFunction{\solution}$ must be polynomially bounded with respect to $\size{\solutionSpace}$.   
\end{enumerate}

\noindent Conditions one and two ensure that the objective function (\cref{eq:objective_generic}) is efficiently computable as classical computation of $\costFunction{\solution}$ is required to compute $\objectiveFunction{}$ and boundedness in $\costFunction{\solution}$ ensures that the number of  measurements required to accurately compute $\objectiveFunction{}$ does not grow exponentially with $\size{\validSolutionSpace}$ \cite{crescenzi_structure_1999}. These conditions constrain the application of QVAs to polynomially bounded (PB) COPs in the non-deterministic polynomial-time optimisation problem (NPO) complexity class (together denoted as NPO-PB) \cite{crescenzi_structure_1999}.

\subsection{Unconstrained Optimisation} \label{sec:unconstrainedOptimisation}

For the case of unconstrained optimisation, the valid solution space $\validSolutionSpace$ is equivalent to $\solutionSpace$. For these COPs a quantum encoding of $\costFunction{\solution}$ is equivalent to the bijective map $\solutionSpace \rightarrow \mathcal{H}$.

\subsubsection{\qaoa{}}

\begin{figure} 
    \centering
	\includegraphics[width=0.8\columnwidth]{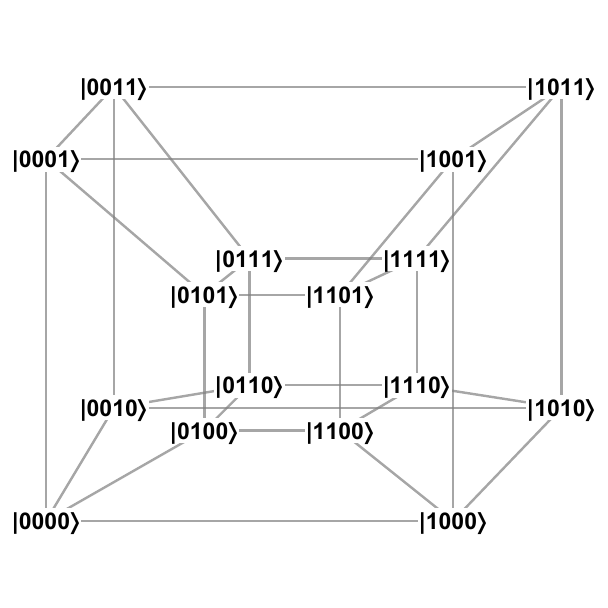}
	\caption{Coupling topology of $W_\text{X}$ in the \qaoa{} for $\size{\solutionSpace} = 16$ ($n$ = 4).}
	\label{fig:hypercube}
\end{figure}

The Quantum Approximate Optimisation Algorithm is comprised of two alternating unitaries. Firstly the phase-shift-unitary

\begin{equation}
    \label{eq:phase_shift_qaoa}
    \U{Q}(\gamma_i) = \exp(-\text{i} \gamma_i \hat{Q})
\end{equation}

\noindent and, secondly, the mixing operator

\begin{equation}
    \label{eq:mixing_qaoa}
    \U{X}(t_i) = \exp(- \text{i} t_i \hat{W}_\text{X}),
\end{equation}

\noindent where $\hat{W}_\text{X} = X^{\otimes N}$ and $X$ is the Pauli-$X$ (or quantum \textsc{NOT}) gate. The mixing operator $\hat{W}_\text{X}$ induces a coupling topology that is equivalent to an $n$-dimension hypercube graph, as shown in \cref{fig:hypercube}.

The initial state $\initialState{QAOA}$ is prepared as an equal superposition over $\mathcal{H}$,

\begin{equation}
    \label{eq:qaoa_initial_state}
    \ket{+} = \frac{1}{\sqrt{n}}\sum_{i = 0}^{n-1}\ket{i}.
\end{equation}  

The final quantum state is then

\begin{equation}
    \label{eq:qaoa}
    \ansatz{QAOA} = \left( \prod_{i=1}^{D} \U{X}(t_i) \U{Q}(\gamma_i)  \right) \ket{+},
\end{equation}

\noindent where $\variationalParameters = \{\gamma_i, t_i \}$ and $\size{\variationalParameters} = 2D$ \cite{farhi_quantum_2014}.

\subsubsection{Extended-QAOA}

A variation of the \qaoa{}, `extended-\qaoa{}' (\extendedQaoa{}), utilises a sequence of phase-shift unitaries,

\begin{equation}
    \label{eq:phase_shift_qaoa_ex}
    \U{Qext}(\gamma_{i}) = \prod_{j=1}^{|\Sigma|} \exp(-\text{i} (\gamma_{i})_{j} \Sigma_{j}),
\end{equation}

\noindent where $\Sigma_j$ are non-identity terms in a Pauli-gate decomposition of $\qualityOperator$ and $\size{\Sigma}$ is the number of non-identity terms \cite{guerreschi_practical_2017}. This increases the number of variational parameters to $\size{\variationalParameters} = (1 + \size{\Sigma})D$ with the intent of achieving a higher degree of convergence to optimal solutions at a lower circuit depth. 

The final state of \extendedQaoa{} is 

\begin{equation}
    \label{eq:qaoa_ex}
   \ansatz{\extendedQaoa{}}= \left( \prod_{i=1}^{D}   \hat{U}_\text{X}(t_i) \hat{U}_\text{Qext}(\gamma_{i,:}) \right) \ket{+},
\end{equation}
\noindent where $\ket{+}$ and $\hat{U}_X$ are defined as in \cref{eq:mixing_qaoa,eq:qaoa_initial_state} and $\variationalParameters = \set{\gamma_{ij}, t_i}$.

\subsection{Constrained Optimisation}\label{sec:constrainedOptimisation}

Constrained optimisation problems seek valid solutions $\validSolution$ from a subset of $\solutionSpace$ as defined by constraints $\solutionConstraints$. Encoding of the solution constraints $\solutionConstraints$ is achieved by restricting the action of the mixing-unitaries $\U{mix}$ and initialising $\initialState{}$ over a subspace of $\mathcal{H}$. 

\subsubsection{\qaoaz{}}

\begin{figure}
    \centering
	\includegraphics[width=0.8\columnwidth]{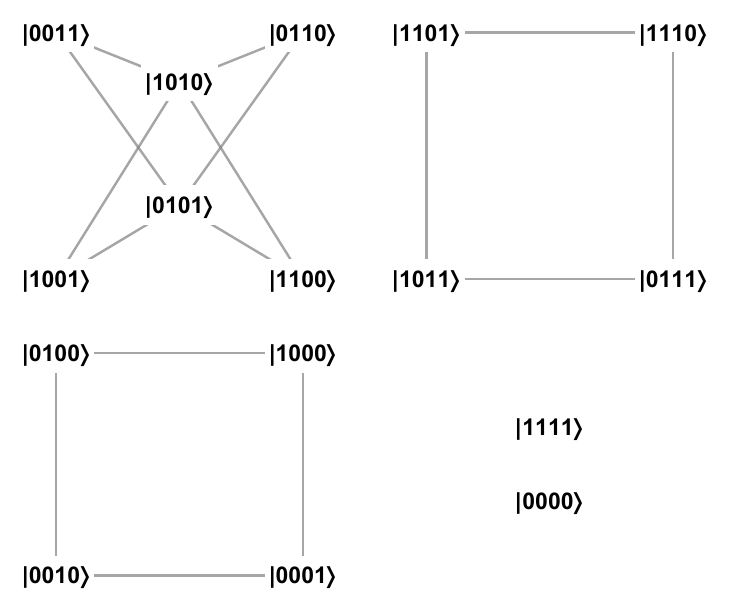}
	\caption{Coupling topology of $\hat{W}$ for the \qaoaz{}  ($n = 4$). Note that $\mathcal{H}$ is partitioned into subgraphs of equal state parity.}
	\label{fig:qaoaz_mixer}
\end{figure}

The Quantum Alternating Operator Ansatz was developed to solve problems for which $\solutionConstraints$ creates a correspondence between $\validSolutionSpace$ and quantum states of equal parity – states with the same number of $\ket{1}$ states. This algorithm consists of the phase-shift-unitary defined in \cref{eq:phase_shift_qaoa}, followed by a sequence of three $\U{mix}$ with mixing operators

\begin{equation}
    \label{eq:parity_terms}
    \begin{aligned}
        & \hat{B}_\text{odd} = \sum_{a \, \text{odd}}^{N-1} X_{a}X_{a+1} + Y_{a}Y_{a+1} \\
        & \hat{B}_{\text{even}} = \sum_{a \, \text{even}}^{N}X_aX_{a+1} + Y_aY_{a+1} \\
        & \hat{B}_\text{last} = 
        \begin{cases}
        X_NX_1 + Y_NY_1, \, \text{odd} \\
        I, \, N \text{even},
        \end{cases}
    \end{aligned}
\end{equation}

\noindent which together form the parity-conserving mixing operator

\begin{equation} \label{eq:qaoaz_mixers}
    \U{parity}(t) = e^{-\text{i} t \hat{B}_\text{last}} e^{-\text{i} t \hat{B}_\text{even}} e^{-\text{i} t \hat{B}_\text{odd}}
\end{equation}

\noindent that mixes probability amplitude between subgraphs of equal parity as illustrated in \cref{fig:qaoaz_mixer}.

By initialising $\initialState{\qaoaz{}}$ in a quantum state that satisfies the parity constraint, probability amplitude is constrained to $\validSolutionSpace$. 

The state evolution of the \qaoaz{} is

\begin{equation}
    \label{eq:qaoaz}
   \ansatz{\qaoaz{}} = \left( \prod_{i=1}^{D}\U{parity}(t_i)\U{Q}(\gamma_i) \right)\initialState{\qaoaz{}}, 
\end{equation}

\noindent where $\initialState{\qaoaz{}}$ is an initial state satisfying the parity constraint \cite{hadfield_quantum_2019}.

\subsubsection{QWOA} \label{sec:qwoa}

The Quantum Walk-assisted Optimisation Algorithm implements $\solutionConstraints$ given the existence of an efficient indexing algorithm for all $\solution \in\ \validSolutionSpace$. Under this condition, the \qwoa{} implements an indexing unitary

\begin{equation}
    \label{eq:qwoa_index}
   \Uindex\ket{i} = 
   \begin{cases}
    \ket{\text{id}_{\solutionConstraints}(i)}, \; \ket{i} \in \ket{\validSolution} \\
    \ket{i}, \; \text{otherwise},
   \end{cases}
\end{equation}

\noindent where $\Uindex$ maps states corresponding to valid solutions $\ket{\validSolution}$ to indexed states $\ket{\text{id}_{\solutionConstraints}(i)}$. By preparing $\initialState{\qwoa{}}$ as an equal superposition over $\ket{\text{id}_{\solutionConstraints}(i)}$ 

\begin{equation}
    \label{eq:qwoa_initial_state}
    \initialState{\qwoa{}} = \frac{1}{\sqrt{\left|\validSolutionSpace\right|}}\sum\limits_{i \in \validSolutionSpace} \ket{i},
\end{equation}

\noindent probability amplitude is restricted to the subspace of indexed states.

The indexing unitary $\Uindex$ and its conjugate unindexing unitary  $\hat{U}_\#$ occur either side of a mixing-unitary that acts on $\ket{\text{id}_{\solutionConstraints}(i)}$:

\begin{equation}
    \label{eq:qwoa_mixer}
        \U{index}(t) = \hat{U}_\# \exp(-i t \hat{W}_\text{QWOA}) \hat{U}_\#^\dag
\end{equation}

\begin{figure}
    \centering
	\includegraphics[width=0.8\columnwidth]{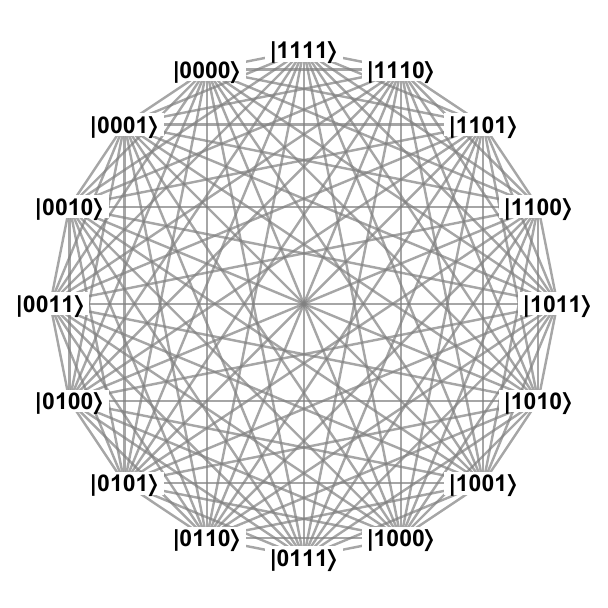}
	\caption{Coupling topology of $\hat{W}$ for the \qwoa{} QWOA ($n = 4$).}
	\label{fig:qwoa_mixer}
\end{figure}

\noindent Where efficiency in the implementation of $\Uindex$ dictates that $\hat{W}_\text{QWOA}$ is circulant. Most commonly, $\hat{W}_\text{QWOA}$ is chosen to be the adjacency matrix of the complete graph as it produces a maximal and unbiased coupling over $\ket{\validSolutionSpace}$ (see \cref{fig:qwoa_mixer}).

The state evolution of the \qwoa{} is

\begin{equation} \label{eq:qwoa}
    \ansatz{\qwoa{}} = \prod_{i = 1}^{D}\U{index}(t_i)\U{Q}(\gamma_i)\initialState{\qwoa{}},
\end{equation}

\noindent where $\variationalParameters = \set{\gamma_i, t_i}$ and there are $\size{\variationalParameters} = 2D$ variational parameters \cite{marsh_combinatorial_2019, marsh_combinatorial_2020}.

\section{Numerical Methods} \label{sec:numerical_methods}

By default, \quop{} presents three approaches by which to compute the action of a phase-shift $\U{phase}$ or mixing-unitary $\U{mix}$. 

As phase-shift unitaries $\U{phase}$ have a diagonal exponent matrix $\hat{O}$, the action of a $\U{phase}(\gamma)$ is efficiently computed by noting that

\begin{equation}
\U{phase}(\gamma)\ket{\psi} = \sum_i^{N-1}e^{-\text{i} \gamma o_i\ket{i}\bra{i}} c_i\ket{i},
\end{equation}

\noindent where $\ket{\psi}$ is an arbitrary quantum state with complex coefficients $c_i$.

For the mixing unitaries $\U{mix}$ , non-diagonal entries in $\hat{W}$, necessitate accurate computation of the action of the matrix exponential. Given a circulant $\hat{W}$, \quop{} takes advantage of the relationship between the eigensystem of circulant matrices and the discrete Fourier transform. The analytical solution for the eigenvalues of a circulant matrix are given by

\begin{equation}
    \lambda_j = w_0 + w_{M-1}\omega^j + w_{M-2} \omega^{2j} + ... + w_1 \omega^{(M-1)j},
\end{equation}
        
\noindent where $M$ is the size of the matrix, $w_{i=0,...,M-1}$ defines the first row of the circulant matrix, $\omega = \exp(\frac{2 \pi \text{i}}{M})$ is a primitive $m^\text{th}$ root of unity and $j=0,..,M-1$. The corresponding eigenvectors,

\begin{equation}
    v_j = \frac{1}{\sqrt{n}} ( \omega^j, \omega^{2j}, ..., \omega^{(M-1)j}),
\end{equation}

\noindent then form the matrix of the discrete Fourier transform. As such, the action of a $\U{mix}$ with a circulant $\hat{W}$ may be implemented as

\begin{equation}
    \U{mix}(t)\ket{\psi} = F^{-1} e^{\text{i} t \Lambda} F\ket{\psi},
\end{equation}

\noindent which is carried out in \quop{} using algorithms provided by the Fastest Fourier Transform in the West (FFTW) library \cite{frigo_fastest_1997, frigo_design_2005}. For the case of sparse mixing operators, \quop{} utilises a variant of the scaling and squaring algorithm, adapted from an implementation previously developed by the authors \cite{matwiejew_qsw_mpi_2021}.  

The above numerical methods support a simulation workflow distinct from gate-based quantum algorithm simulation packages. For instance, efficient gate-based simulation of the complete wavefunction can be achieved by combining one and two-qubit gates to reduce the total number of required matrix multiplications \cite{broughton_tensorflow_2020}. \cite{broughton_tensorflow_2020}. Alternatively, tensor-network based approximations reduce the computational cost by disregarding long-range interactions or qubit couplings, in addition to forming an efficient decomposition of the quantum circuit as a sequence of tensor products \cite{mccaskey_xacc_2020}. 

 Gate based simulation efficiency is highly dependent on the structure of the $\U{mix}$ and $\U{phase}$ matrix exponents. In general, $\U{mix}$ and $\U{phase}$ are approximated as per the quantum Hamiltonian Simulation Algorithm \cite{tensor_github}, which is based on a Trotter-Suzuki decomposition of the matrix exponential \cite{nielsen_quantum_2010}. Such representations are computationally efficient given $\U{mix}$ with sparse matrix exponents expressed in the Pauli basis \cite{nielsen_quantum_2010}. However, accurate simulation of arbitrary mixing operators is generally not possible as the computational cost of simulation is proportional to the length of the quantum circuit \cite{nielsen_quantum_2010}. Efficient gate-base representations can also be hard to realise for highly entangling quantum algorithms - which offer some of the best examples of quantum advantage. For instance, one such algorithm, the Quantum Fourier Transform, offers an exponential advantage over its classical counterpart \cite{hales_fourier}. 
 
 For these reasons, the numerical methods provided with \quop{} focus on providing QVA simulations to double-precision accuracy in a manner that is agnostic to any specific implementation. Together, \quop{}, and other gate-based simulation packages provide for different avenues of investigation. The former enables research into the limiting characteristics of QVAs, and the latter supports the investigation of gate-based realisations of QVAs.

The choice of initial values for the variational parameters $\variationalParameters{}$ and the accompanying classical optimisation algorithm is an active area of research \cite{zhou_quantum_2020}. By default, \quop{} uses the Broyden-Fletcher-Goldfarb-Shanno (BFGS) algorithm \cite{nocedal_numerical_2006} provided by SciPy via its \verb|minimize| function \cite{jones_scipy_2001} as the authors have found it to behave reliably across a wide variety of QVAs (see \cref{sec:performence}). Users are able to adjust the parameters and optimisation algorithms used by the \verb|minimize| function or opt to use algorithms provided by the NLOpt package \cite{johnson_nlopt_nodate} through an included `SciPy-like' interface \cite{steinberg_revrand_nodate}. 

\section{Parallelisation Schemes}\label{sec:parallel}

Parallelisation in \quop{} is implemented using the Message Passing Interface (MPI) standard. In general terms, MPI supports a distributed-memory model of parallel computing in which concurrent instances of the same program operate within isolated memory and namespaces, communicating with each other using `message-passing' directives. As opposed to shared-memory parallel frameworks, this allows for the use of large scale distributed computers (i.e. supercomputers). 

A group of program copies (MPI \emph{processes}) that are capable of MPI communication form an MPI \emph{communicator}. Within an MPI communicator, each process is identified by a sequential \emph{rank} ID that ranges from $0$ to $\mpiSize - 1$, where $\mpiSize$ is the total number of MPI processes in the communicator. Communicator subsets (\emph{sub-communicators}) can be created and assigned to sub-tasks. Note that, while an MPI process is also commonly referred to as a \emph{node}, in this work \emph{node} refers only to a compute-node in a computational cluster. Depending on user-controlled settings, \quop{} operates on the default global MPI communicator or a variable configuration of MPI sub-communicators.

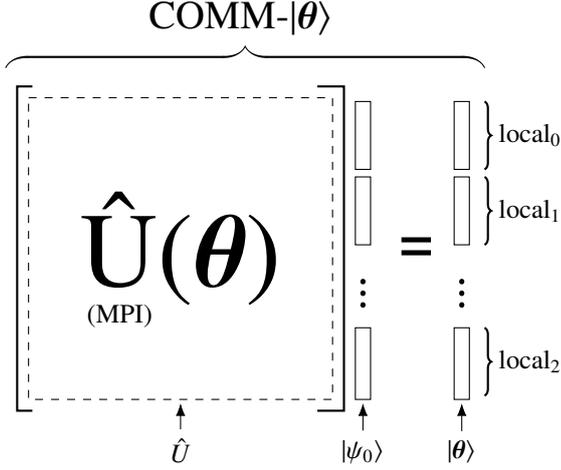
\begin{figure}
\centering
    \input{mpi-theta.tikz}
\caption{The partitioning of $\initialState{}$ and $\ansatz{}$ over an MPI communicator of $\mpiSize = l$ where $\mpiLocalElements_i$ denotes the number of vector elements stored at \mpiRank{} $i$. The ansatz unitary $\U{}$ is implemented as a sequence of MPI parallelised subroutines that recieve the distributed state vector and $\theta_i$ as inputs and return $\ansatz{}$ as an identically distributed state vector.}
\label{fig:mpi-theta}
\end{figure}

The primary \quop{} parallelisation scheme $\mpiAlg{}$ is illustrated in \cref{fig:mpi-theta}. The initial state $\initialState{}$, evolved state $\ansatz{}$ and observable values $\qualityVector{}$ are distributed over an MPI communicator with each $\mpiRank{}$ containing a partition of sequential elements. The global position of a vector partition is thus specified by two variables; the number of vector elements in the local partition 

\begin{equation} \label{eq:local_i}
	\mpiLocalElements_i
\end{equation} 

\noindent and the vector element index offset

\begin{equation} \label{eq:offset}
	\text{local\_i\_offset} = \sum_{m = 0}^{\mpiRank - 1} \mpiLocalElements_m.
\end{equation}

\noindent Unitary evolution is carried out using MPI-parallelised subroutines that act on the local vector partitions.  

At run-time, \quop{} attempts to partition the global vectors equally over each rank in $\mpiAlg{}$ while satisfying any partitioning constraints associated with external libraries such as FFTW. If a process receives zero vector elements, they are excluded from $\mpiAlg{}$ and the vector partitioning is recalculated. State evolution and calculation of $\objectiveFunction{}$ is then carried out over $\mpiAlg$ in parallel. 

With each evaluation of $\ansatz{}$, $\objectiveFunction{}$ is sent to rank $0$ of $\mpiAlg{}$ where it is received by the optimisation algorithm. The adjusted $\variationalParameters$ are then broadcast to all nodes in $\mpiAlg{}$, and the cycle repeats until the optimisation algorithm terminates. The distributed $\ansatzFinal{}$ may then be written to disk using parallel HDF5 or gathered at the root $\mpiRank$. 

\begin{figure}
\centering
    \input{mpi-jac.tikz}
\caption{Parallel computation of the objective function gradient $\gradient{}$ where each box represents an MPI sub-communicator implementing an instance of $\mpiAlg{}$. The objective function $\objectiveFunction{}$ is calculated by the $\mpiAlg{}$ containing the MPI process with $\mpiRank = 0$ in the global MPI communicator and the partial derivatives by sub-communicators $1$ through to $m$ where $m \leq \size{\variationalParameters{}}$. }
\label{fig:mpi-gradient}
\end{figure}

In the case of optimisation algorithms that make use of gradient information, the user may choose to calculate the objective function gradient $\gradient{}$ in parallel. As shown in \cref{fig:mpi-gradient}, the global communicator is split into $m + 1$ MPI sub-communicators of  which $m$ are assigned the task of approximating the partial derivative of $\objectiveFunction{}$ (via forward or central differences) for a subset of $\variationalParameters{}$. The number of created sub-communicators depends on $\size{\variationalParameters{}}$, the number of compute nodes in the global communicator and a user-definable `\verb|parallel|' attribute (see \cref{tab:ansatz}):

\begin{enumerate}
 \item If $\size{\variationalParameters{}} < \text{nodes} + 1$ and \verb|parallel = `jacobian'|: create sub-communicators consisting of multiple nodes.
 \item If $\size{\variationalParameters} > \text{nodes} + 1$ and \verb|parallel = `jacobian'|: create multiple sub-communicators within each compute node.
 \item If \verb|parallel = `jacobian_local'|: create one sub-communicator per compute node.
\end{enumerate}

\noindent The default behaviour of \quop{} is parallelisation of $\mpiAlg$ exclusively (\verb|parallel = `global'|) . Together, $\mpiAlg$ and $\mpiJac$ allow the user to specify an MPI process configuration that is optimal for their hardware and simulation scale. 

\section{Package Overview} \label{sec:overview}

\quop{} is a \python{} module that provides an object-orientated approach to QVA design and parallel simulation. Foremost, it presents an approachable workflow in which users can write efficient and scalable quantum simulations without requiring prerequisite knowledge of complied programming languages or parallel programming techniques. This is underpinned by flexible class structures and parallelisation schemes that have been designed to streamline the integration of additional parallel simulation methods.

\begin{figure}[bh]
    \centering
    \includegraphics[width=0.85\columnwidth]{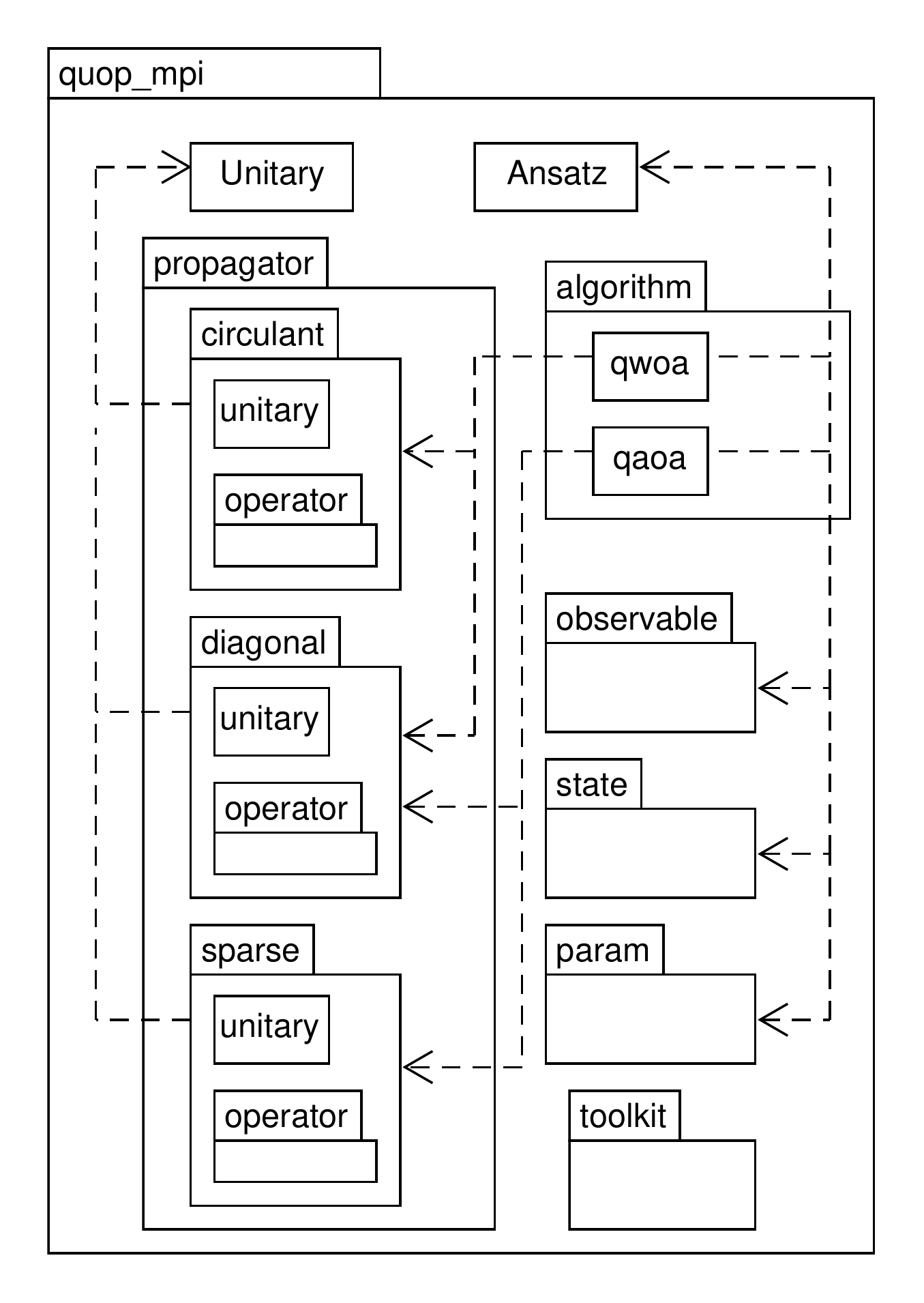}
	\caption{The user-level package structure of the \quop{} Python module.}
	\label{fig:package_diagram}
\end{figure}

\begin{figure}[p!]
    \centering
    \includegraphics[width=0.9\columnwidth]{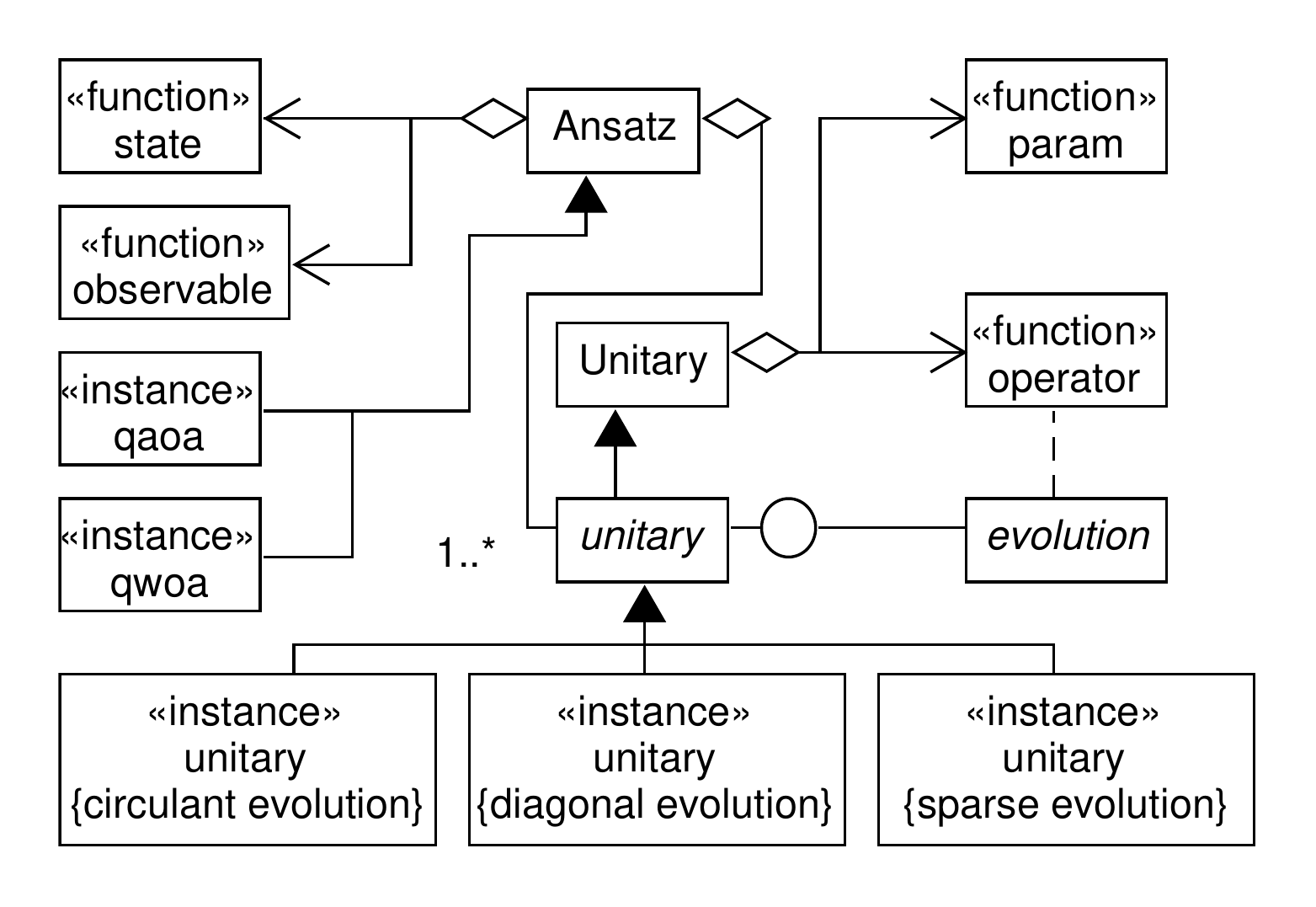}
	\cprotect\caption{\quop{} class structure. The \verb|Unitary| class provides an interface to a state \emph{evolution} method that is compatible with the $\mpiAlg$ parallelisation scheme (see \cref{sec:parallel}) that, when aggregated with an \verb|operator| and \verb|param| function, implements the action of a $\U{phase}$ or $\U{mix}$ on $\ket{\psi}$. The \verb|Ansatz| class coordinates evaluation and minimisation of $\objectiveFunction{}$ when aggregated with a list of \verb|unitary| instances, a \verb|state| function and an \verb|observable| function (see \cref{tab:functions}). \quop{} includes \verb|unitary| classes for circulant, diagonal and sparse matrix exponents (see \cref{sec:numerical_methods}), and two \verb|Ansatz| instances, \verb|qaoa| and \verb|qwoa|, which implement simulation of the \qaoa{} and \qwoa{}.} 
	\label{fig:class_diagram}
\end{figure}

\input{unitary_class_table.tex}

\input{functions_table}

\input{ansatz_class_table.tex}

The user-level structure of \quop{} is shown in \cref{fig:package_diagram}. The package is centred around the \verb|Ansatz| and \verb|Unitary| `template' classes. The \verb|Ansatz| class manages the parallelisation scheme, definition, execution of the QVA and the recording of simulation results for a specific QVA. The \verb|Unitary| class provides a scaffolding with which parallel algorithms for the computation of the action of $\U{phase}$ or $\U{mix}$ on $\ket{\psi}$ are integrated with \quop{}.  Overviews of the key methods of the \verb|Ansatz| class and the user-implemented methods required to integrate a state evolution method into \quop{} via the \verb|Unitary| class are given in \cref{tab:ansatz,tab:unitary} respectively.

As shown in \cref{fig:class_diagram,tab:unitary}, a \quop{} compatible method for simulating the action of a $\U{phase}$ or $\U{mix}$ on $\ket{\psi}$ is implemented through the creation of an \verb|Unitary| subclass (\verb|unitary|) which defines methods responsible for determination of the $\mpiAlg$ parallel partitioning scheme (see \cref{fig:mpi-theta}), computation of the action of the unitary and management of the ancilla requirements of any external subroutines. 

The \verb|propagator| submodule contains predefined \verb|unitary| classes together with an \verb|operator| submodule containing functions for the generation of compatible matrix exponents $\hat{O}$ or $\hat{W}$. Three \verb|unitary| classes are included as part of the \verb|diagonal|, \verb|sparse| and \verb|circulant| submodules, which simulate $\U{phase}$, $\U{mix}$ with sparse matrix exponents and $\U{mix}$ circulant matrix exponents (see \cref{sec:numerical_methods}).  

\FloatBarrier

Submodules \verb|state|, \verb|param| and \verb|observable| provide functions that, when passed to the \verb|Ansatz| class, define $\initialState{}$, $\variationalParameters_0$ and $\qualityVector$ for a particular QVA. 

Two predefined QVAs are included in the \verb|algorithm| submodule, \verb|qwoa| and \verb|qaoa|. These \verb|Ansatz| subclasses implement the \qwoa{} and \qaoa{} respectively. The \verb|toolkit| submodule provides convenience functions to assist in constructing matrix operators and quantum states involving the tensor product of Pauli matrices and bit-string qubit states.

Finally, \quop{} is highly extensible through its support for user-defined functions for the generation of the matrix exponents, $\variationalParameters_0$, $\qualityVector$ and $\initialState{}$, as described in \cref{tab:functions}.

\begin{figure*}[]
\centering
\begin{tcolorbox}[enhanced, fontupper=\scriptsize, boxsep=5pt, left = 10pt, right=5pt, top=2pt, bottom=2pt, width=\linewidth, colframe=black!100!white!20, colback=white, coltitle = black, title = {\begin{footnotesize}\quop{}/examples/maxcut/maxtcut.py\end{footnotesize}}]
\setlength{\columnsep}{1.2cm}
\setlength{\columnseprule}{0.2pt}
\renewcommand\columnseprulecolor{\hspace{-0.05cm}}
\begin{multicols}{2}
\begin{lstlisting}[language=Python, numbers=left,upquote=true, numbersep=5pt, numberstyle=\tiny\color{gray}]
from quop_mpi.algorithm import qaoa
from quop_mpi import observable
from quop_mpi.toolkit import I, Z
import networkx as nx

graph = nx.circular_ladder_graph(4)

n_qubits = graph.number_of_nodes()
system_size = 2 ** n_qubits

def maxcut_qualities(graph, n_qubits):
    C = 0
    for edge in graph.edges():
        C += 0.5*(I(n_qubits) + \
	(Z(edge[0], n_qubits) @ Z(edge[1], n_qubits)))
    return C.diagonal()

alg = qaoa(system_size)

alg.set_qualities(observable.serial,
	{"function": maxcut_qualities,
	"args": [graph, n_qubits]})

alg.set_depth(2)
alg.execute()
alg.print_optimiser_result()
alg.save("maxcut", "depth 2", "w")
\end{lstlisting}
\end{multicols}
\end{tcolorbox}
\caption{Example 1: Simulation of the \qaoa{} applied to the max-cut problem.}
\customlabel{ex-1}{1}
\end{figure*}

Within this structure, \quop{} thus presents several levels of usage:

\begin{enumerate}
	\item Simulation of the \qwoa{} or \qaoa{} using the \verb|qwoa| or \verb|qaoa| classes with user defined $\qualityVector$.
	\item Simulation of the \qwoa{} or \qaoa{} using user-defined parallel generation of $\qualityVector$.
	\item Design and simulation of a QVA with the \verb|Ansatz| class with included or user defined functions specifying matrix exponents, $\variationalParameters_0$, $\qualityVector$ and $\initialState{}$.
	\item Integration of additional state evolution methods for $\U{mix}$ or $\U{phase}$ through the creation of \verb|Unitary| subclasses.
\end{enumerate}

\begin{figure}
	\centering
	\footnotesize{\quop{}/examples/maxcut\_plots.py}

	\includegraphics[width=0.65\columnwidth]{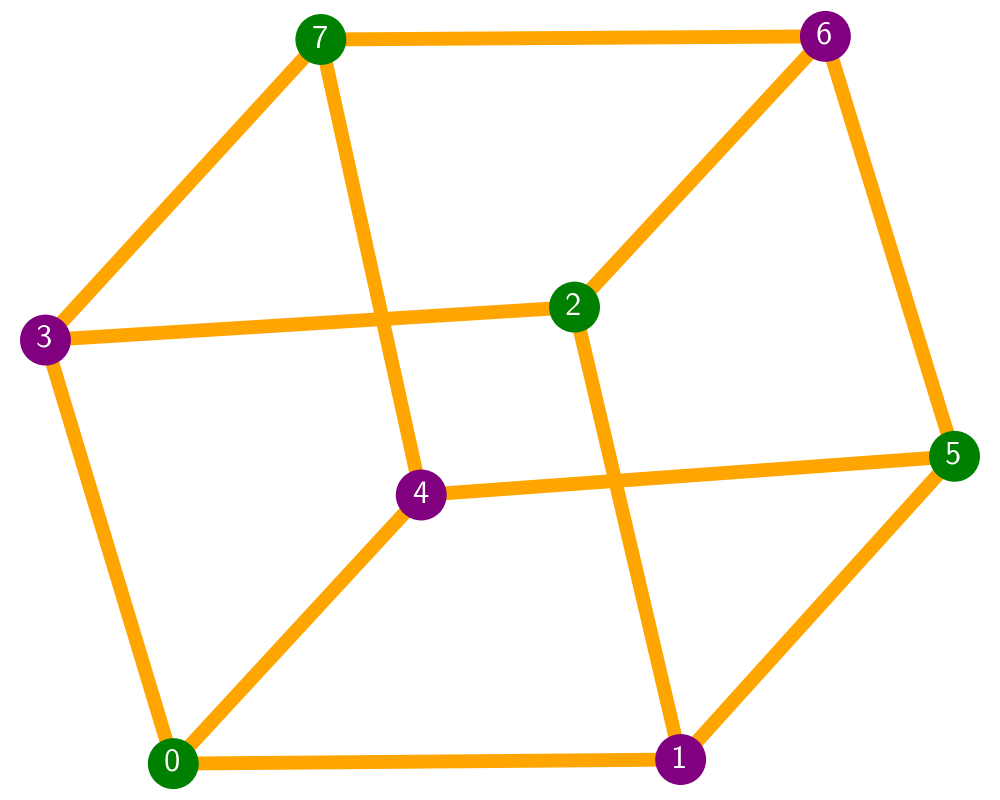}

	\caption{Graph for which the max-cut problem is solved in Examples \ref{ex-1} and \ref{ex-2}. The most probable solution for $\ansatz{\qaoa}$ and $\ansatz{\extendedQaoa{}}$  ($\optimalSolution = (0,1,0,1,1,0,1,0)$) is shown by vertex colouring with purple (darker) indicating a $0$ and green (lighter) indicating a $1$. This partitioning corresponds to the optimal solution for which $C(\optimalSolution) = {q_{90}} = 0$.}
	\label{fig:max-cut-solution}
\end{figure}

\begin{figure}
	\centering
	\footnotesize{\quop{}/examples/maxcut\_plots.py}

	\includegraphics[width=0.75\columnwidth]{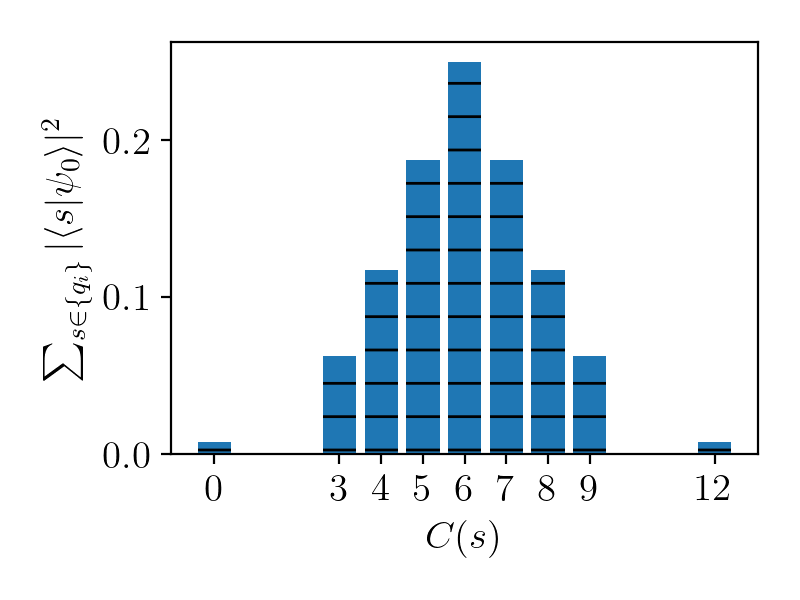}

	\caption{The initial solution probability distribution for the max-cut problem simulated in Examples \ref{ex-1} and \ref{ex-2}.}
    \label{fig:maxcut-qaoa-initial}
\end{figure}

\section{Usage Examples} \label{sec:usage}

The following introduces typical \quop{} usage by simulating the \qaoa{}, \extendedQaoa{}, \qaoaz{} and \qwoa{} as applied to the max-cut and portfolio-re-balancing optimisation problems. 

\subsection{The max-cut problem.}

The max-cut problem seeks to partition the vertices of a graph such that a maximum number of neighbouring nodes are assigned to two disjoint sets \cite{farhi_quantum_2014}. A quantum encoding of the max-cut problem is a bijective mapping of the vertices of a graph $G$ to $n$ qubits, with the set membership indicated by the corresponding qubit state. For example, a two vertex graph with vertices $\set{0,1}$ has a solution space that is completely represented by an equal superposition over a two-qubit system: $\set{\set{0,1}} \rightarrow \ket{00}$, $\set{\set{0}, \set{1}} \rightarrow \ket{01}$, $\set{\set{0},\set{1}} \rightarrow \ket{10}$  and $\set{\set{0,1}} \rightarrow \ket{11}$.

The cost function is then implemented as

\begin{equation}
	\label{eq:maxcut-cost}
    \costFunction{\solution} = \sum_{E(i,j)\in G} \frac{1}{2}\left( \mathbb{I} + Z_iZ_j\right),
\end{equation}

\noindent where $Z_i$ is a Pauli Z gate acting on the $i^\text{th}$ qubit, $E(i,j)$ is an edge in $G$ connecting vertex $i$ to vertex $j$, and $Z_iZ_j$ has eigenvalue $1$ if qubits $i$ and $j$ are in the same state or $-1$ otherwise.

\subsubsection{\qaoa{}}\label{sec:example_qaoa}

In Example \ref{ex-1} the \qaoa{} is applied to the max-cut problem for the graph shown in \cref{fig:max-cut-solution}. The predefined \verb|Ansatz| subclass \verb|qaoa| forms the basis of the simulation.

To generate the graph we use the external module \verb|networkx| (Line 6). On Lines 11 to 16, the cost function is defined. By using the \verb|I| and \verb|Z| functions from the \verb|toolkit| submodule, we are able to directly implement \cref{eq:maxcut-cost}. The matrices computed on Lines 14 and 15 are in a SciPy sparse matrix format.

\begin{figure*}[t!]
\begin{tcolorbox}[enhanced, fontupper=\scriptsize, boxsep=5pt, left = 10pt, right=5pt, top=2pt, bottom=2pt, width=\linewidth, colframe=black!100!white!20, colback=white, coltitle = black, title = {\begin{footnotesize}\quop{}/examples/max-cut\_extended/maxcut\_extended.py\end{footnotesize}}]
\setlength{\columnsep}{2.35cm}
\setlength{\columnseprule}{0.2pt}
\renewcommand\columnseprulecolor{\hspace{1.2cm}}
\begin{multicols}{2}
\begin{lstlisting}[language=Python,numbers=left,upquote=true, numbersep=8pt, numberstyle=\tiny\color{gray}]
from quop_mpi import Ansatz
from quop_mpi.propagator import diagonal, sparse
from quop_mpi.observable import serial
from quop_mpi.param.rand import uniform
from quop_mpi.toolkit import I, Z
import numpy as np
import networkx as nx

graph = nx.circular_ladder_graph(4)

n_qubits = graph.number_of_nodes()
n_edges = 2 * graph.number_of_edges()

system_size = 2 ** n_qubits

def maxcut_terms(graph, n_qubits):
    terms = []
    for edge in graph.edges:
        term = 0.5*(I(n_qubits) + Z(edge[0], n_qubits) @ \
		Z(edge[1], n_qubits))
        terms.append(term.diagonal())
    return terms

def maxcut_qualities(terms):
    return np.sum(terms, axis=0)

computed_terms = maxcut_terms(graph, n_qubits)

UQ = diagonal.unitary(
    diagonal.operator.serial,
    operator_kwargs={"function": maxcut_terms,
		    "args": [graph, n_qubits]},
    unitary_n_params=n_edges,
    parameter_function=uniform)

UW = sparse.unitary(sparse.operator.hypercube, 
	            parameter_function=uniform)

alg = Ansatz(system_size)

alg.set_unitaries([UQ, UW])

alg.set_observables(serial,
	{"function": maxcut_qualities,
	"args": [computed_terms]})

alg.execute()
alg.print_optimiser_result()
alg.save("maxcut_extended", "depth 1", "w")
\end{lstlisting}
\end{multicols}
\end{tcolorbox}
\caption*{Example 2: Simulation of the \extendedQaoa{} applied to the max-cut problem.}
\customlabel{ex-2}{2}
\end{figure*}

\begin{figure}[t]
	\centering
	\vspace{0.1cm}
	\footnotesize{\quop{}/examples/maxcut\_plots.py}

    \includegraphics[width=0.75\columnwidth]{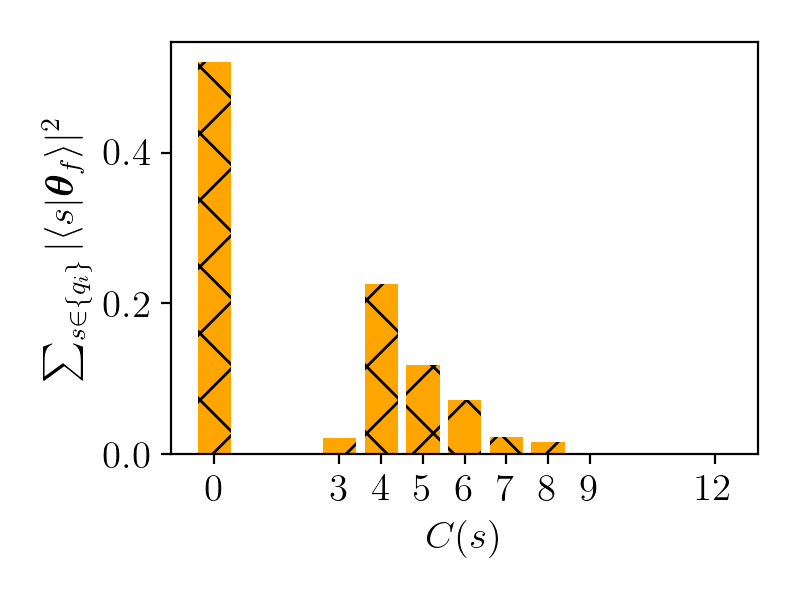}

	\caption{Solution quality probability distribution of $\ansatz{\qaoa}$ as simulated in Example \ref{ex-1}.}
    \label{fig:maxcut-qaoa-final}
\end{figure}

Lines 18 to 27 demonstrate standard use of the \verb|qaoa| class. An instance of the class is instantiated for \verb|system_size|$=N$. Next, the $\qualityVector$ is defined via the \verb|set_qualities| method. For this, we pass the \verb|serial| \verb|observable| function along with a dictionary of its keyword arguments. The \verb|serial| function assists with memory-efficient simulation given a serial \verb|observable| function by calling the function at the root MPI process and distributing its output over $\mpiAlg$. The ansatz depth ($D=2$) is then defined via the \verb|set_depth| method.

Now that the \verb|qaoa| instance is fully specified, simulation of the algorithm (as defined in \cref{eq:qaoa}) proceeds via the \verb|execute| method. By calling \verb|execute| without specifying $\variationalParameters{}$ we choose to use default \verb|param| functions which generate $\theta_i$ from a uniform distribution over $(0 \pi, 2\pi]$.

Finally, the optimiser result is displayed using the \verb|print_optimiser_result| method and the simulation results are saved to the HDF5 file `maxcut.h5' under the `depth 2' group. As compared to a starting expectation value of $7.43$, the final value of the objective function (\verb|fun|) is approximately $2.30$ with variational parameters (\verb|x|) 3.59, 6.88, 3.95 and 5.92. 

\begin{figure}[t!]
\begin{tcolorbox}[enhanced, fontupper=\scriptsize, boxsep=5pt, left = 10pt, right=10pt, top=2pt, bottom=2pt, width=\columnwidth, colframe=black!100!white!20, colback=white, coltitle = black, title = {\begin{footnotesize}\quop{}/examples/maxcut/maxcut\_parallel\_qualities.py\end{footnotesize}}]
\begin{lstlisting}[language=Python,numbers=left,upquote=true, numbersep=5pt, numberstyle=\tiny\color{gray},firstline=1,lastline= 17]
def maxcut(local_i, local_i_offset, graph=None):
        
    n_qubits = graph.number_of_nodes()
    n_edges = graph.number_of_edges()
    
    q = np.full(local_i, n_edges, dtype = np.float64)

    start = local_i
    end = local_i + local_i_offset

    for i in range(start, end):
        bs = np.binary_repr(i, width = n_qubits)
        for edge in graph.edges:
            if bs [edge[0]] != bs [edge[1]]:
                q[i - local_i_offset] -= 1

    return q
\end{lstlisting}
\end{tcolorbox}
\caption*{Example 3: User-defined quality function for the max-cut problem.}
\customlabel{ex-3}{3}
\end{figure}

\cref{fig:maxcut-qaoa-initial,fig:maxcut-qaoa-final}, illustrates the initial and final probability distributions with respect to unique values of $q_i$. After application of the \qaoa{} to the initial superposition, probability density is concentrated at high-quality solutions with the optimal solution ($q_{90} = 0$) having the highest probability of measurement.

\begin{figure*}
\begin{tcolorbox}[enhanced, fontupper=\scriptsize, boxsep=5pt, left = 10pt, right=5pt, top=2pt, bottom=2pt, width=\linewidth, colframe=black!100!white!20, colback=white, coltitle = black, title = {\begin{footnotesize}\quop{}/examples/portfolio\_rebalancing/qwoa\_portfolio.py\end{footnotesize}}]
\setlength{\columnsep}{2cm}
\setlength{\columnseprule}{0.2pt}
\renewcommand\columnseprulecolor{\hspace{1cm}}
\begin{multicols}{2}
\begin{lstlisting}[language=Python,numbers=left,upquote=true, numbersep=5pt, numberstyle=\tiny\color{gray}]
from quop_mpi.algorithm import qwoa
from quop_mpi import observable
import pandas as pd

qualities_df = pd.read_csv('qwoa_qualities.csv')
qualities = qualities_df.values[:,1]

system_size = len(qualities)

alg = qwoa(system_size)

alg.set_qualities(
        observable.array,
        {'array':qualities})

alg.set_log(
        'qwoa_portfolio_log',
        'qwoa',
        action = 'w')

alg.benchmark(
        range(1,6),
        3,
        param_persist = True,
        filename = 'qwoa_portfolio',
        save_action = 'w')
\end{lstlisting}
\end{multicols}
\end{tcolorbox}
\caption*{Example 4: Simulation of the \qwoa{} applied to the portfolio re-balancing problem.}
\customlabel{ex-4}{4} 
\end{figure*}

\subsubsection{Extended-QAOA}\label{sec:example_exqaoa}

Having demonstrated the effectiveness of the \qaoa{} in finding high-quality max-cut solutions, we will now explore the application of the \extendedQaoa{} to the same task. Example \ref{ex-2} demonstrates the implementation of \extendedQaoa{} using the \verb|Ansatz| class. As in Example \ref{ex-1}, the graph and its adjacency matrix are generated using \verb|networkx|.

The functions needed to implement \cref{eq:phase_shift_qaoa_ex} and \cref{eq:mixing_qaoa} are defined from Lines 16 to 22 and Lines 24 to 25 respectively. The first of these, \verb|maxcut_terms|, returns an array of the summation terms in \cref{eq:maxcut-cost} with which the \verb|maxcut_qualities| function returns $\qualityVector{}$. 

A two-step calculation of the solution qualities is chosen as the \extendedQaoa{} phase-shift operator associates a $\theta_i$ with each term of a Pauli-matrix decomposition of $\qualityVector$. The phase-shift-unitary is implemented using the \verb|propagator| submodule \verb|diagonal|. An instance of the \verb|diagonal| submodule \verb|unitary| class (\verb|UQ|) is defined on Lines 29 to 34. The first argument specifies the \verb|operator| function responsible for generating the Pauli-matrix terms $\Sigma_j$. The second specifies a dictionary of user-defined keyword arguments for the \verb|operator| function. The third argument specifies the number of $\theta_i$ associated with \verb|UQ| and, finally, the fourth argument specifies the \verb|param| function used to initialise the unitary's variational parameters. The \verb|param| function generates $\theta_i$ as described in \cref{sec:example_qaoa}.

The \verb|operator| function \verb|diagonal.serial| executes the serial \verb|maxcut_terms| function at the root MPI process and distributes the array of Pauli-matrix terms over $\mpiAlg$. The \verb|unitary_n_params| keyword argument describes the number of operator terms returned by \verb|diagonal.serial|, which are mapped to a sequence of $\U{phase}$ unitaries each parameterised by a unique $\theta_i$.

Definition of the mixing-unitary \verb|UW| occurs on Lines 36 to 37. As with \verb|UQ|, the first argument specifies the \verb|operator| function and the \verb|parameter_function| argument specifies the \verb|param| function. The \verb|operator| function \verb|sparse.operator.hypercube| generates a parallel-partitioned instance of the hypercube mixing operator (see \cref{eq:qwoa_mixer}).

On Line 41 the defined unitaries \verb|UQ| and \verb|UW| are passed to an instance of the \verb|Ansatz| class via the \verb|set_unitaries| method.
The objective function is then defined by passing the \verb|maxcut_qualities| function to the \verb|set_observables| method.

\begin{figure}[H]
	\centering
\footnotesize{\quop{}/examples/maxcut\_extended/maxcut\_extended\_plots.py}
		\includegraphics[width=0.75\columnwidth]{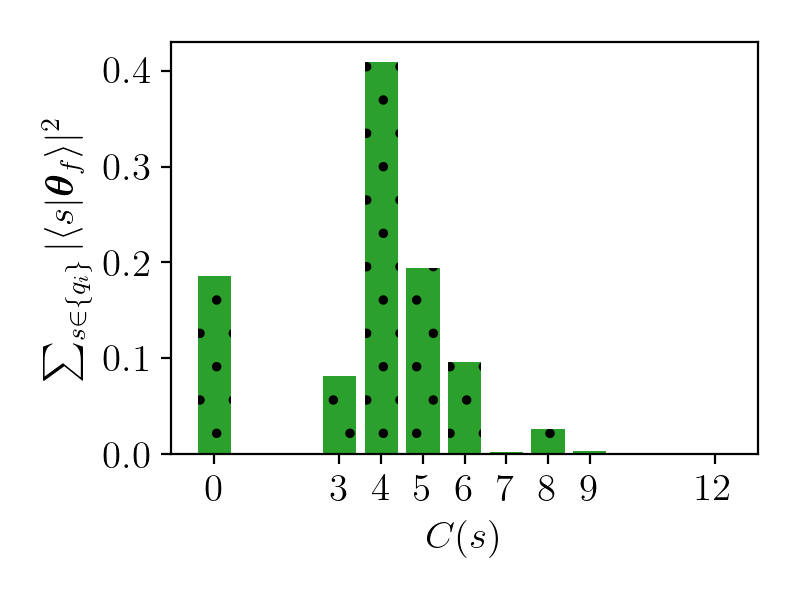}
	\caption{Final probability distribution of the max-cut solutions  following the execution of the \extendedQaoa{} defined in Example (\ref{ex-2}).}
	\label{fig:max-cut-extended-qaoa}
\end{figure}

The \extendedQaoa{} simulation is then executed on Line 47. As $D$ has not been specified via the \verb|set_depth| method, the algorithm is simulated with the default ansatz depth of $D = 1$ (see \cref{fig:max-cut-extended-qaoa}).

\subsubsection{Parallel Computation of the Cost Function}\label{sec:user_fun}

As a corollary of condition two in \cref{sec:CO} it will generally be the case that computation of any particular $C(s)$ is independent of the rest of the cost function values. In such instances, the generation of $\qualityVector{}$ is an embarrassingly parallel problem, and, as such, users are encouraged to implement parallel quality functions. A parallel \verb|observable| function for the max-cut problem is shown in Example \ref{ex-3} as per the requirements described in \cref{tab:functions}. At run-time, this function is called by each rank in $\mpiAlg$ to generate the $q_i$ specific to the local vector partitions. 

\begin{figure*}[t!]
\begin{tcolorbox}[enhanced, fontupper=\scriptsize, boxsep=5pt, left = 6pt, right=5pt, top=2pt, bottom=2pt, width=\linewidth, colframe=black!100!white!20, colback=white, coltitle = black, title = {\begin{footnotesize}\quop{}/examples/portfolio\_rebalancing/qaoaz\_portfolio.py\end{footnotesize}}]
\setlength{\columnsep}{1.5cm}
\setlength{\columnseprule}{0.2pt}
\renewcommand\columnseprulecolor{\hspace{0.5cm}}
\begin{multicols}{2}
\begin{lstlisting}[language=Python, numbers=left,upquote=true, numbersep=4pt, numberstyle=\tiny\color{gray}]
from quop_mpi import Ansatz, observable, state, param
from quop_mpi.propagator import diagonal, sparse
from quop_mpi.toolkit import kron, kron_power
from quop_mpi.toolkit import string, X, Y
from qaoaz_qualities import qaoaz_portfolio
from numpy import sqrt

def parity_ring(i, j, n_qubits):
    parity = X(i, n_qubits) @ X(j, n_qubits) \
            + Y(i, n_qubits) @ Y(j, n_qubits)
    return parity

def parity_mixer(qubits, n_qubits):

    odd = 0
    even = 0

    n_subset = len(qubits)

    for i in range(n_subset - 1):

        if (i % 2 != 0):
            odd += parity_ring(qubits[i],
                    qubits[(i + 1) % n_subset],
                    n_qubits)

        elif i % 2 == 0:
            even += parity_ring(qubits[i],
                    qubits[(i + 1) % n_subset],
                    n_qubits)

    mixer = [odd, even]

    if len(qubits) % 2 != 0:
        last = parity_ring(qubits[-1],
                qubits[1],
                n_qubits)

        mixer.append(last)

    return mixer

def mixer(n_qubits):
    short_qubits = [i for i in range(0, n_qubits, 2)]
    long_qubits = [i for i in range(1,n_qubits, 2)]
    short_mixer = parity_mixer(short_qubits, n_qubits)
    long_mixer = parity_mixer(long_qubits, n_qubits)
    return short_mixer + long_mixer

def parity_state(n_qubits, D):
    M = n_qubits//2
    term_1 = kron_power(string('01'), D)
    term_2 = kron_power(
        1/sqrt(2) * (string('11') + string('00')),
        M-D)
    state = kron([term_1, term_2])
    return state

n_qubits = 8
system_size = 2**n_qubits

UQ = diagonal.unitary(
        qaoaz_portfolio,
        parameter_function = param.rand.uniform)

UW = sparse.unitary(
        sparse.operator.serial,
        operator_kwargs = {
            'function':mixer,
            'args':[n_qubits]},
        parameter_function = param.rand.uniform)

alg = Ansatz(system_size)

alg.set_unitaries([UQ, UW])

alg.set_initial_state(
        state.serial,
        {'function':parity_state,
            'args':[n_qubits, 2]})

alg.set_observables(0)

alg.set_log(
        'qaoaz_portfolio_log',
        'qaoaz',
        action = 'w')

alg.benchmark(
        range(1,6),
        3,
        param_persist = True,
        filename = 'qaoaz_portfolio',
        save_action = 'w')
\end{lstlisting}
\end{multicols}
\end{tcolorbox}
\caption*{Example 5: Simulation of the \qaoaz{} applied to the portfolio re-balancing problem.}
\customlabel{ex-5}{5}
\end{figure*}

\subsection{Portfolio Re-balancing}

To explore the case of constrained optimisation using the \qwoa{} and the \qaoaz{} we will consider the problem of portfolio re-balancing. For each asset in a portfolio of size $M$, an investor must choose one of the following positions:

\begin{enumerate}
	\item Short position: buying and selling an asset with the expectation that it will drop in value.
	\item Long position: buying and holding the asset with the expectation that it will rise in value.
	\item No position: taking neither the long or short position.  
\end{enumerate}

A quantum encoding of the possible solutions uses two qubits per asset.

\begin{enumerate}
	\item $\ket{01} \rightarrow \text{short position}$
	\item $\ket{10} \rightarrow \text{long position}$
	\item $\ket{00}$ or $\ket{11} \rightarrow \text{no position}$
\end{enumerate}

The discrete mean-variance Markowitz model provides a means of evaluating the quality associated with a given combination of positions. It can be expressed through minimisation of the cost function,

\begin{equation}
\costFunction{\solution} = \omega \sum_{i,j = 1}^{M} \sigma_{ij}Z_iZ_j - (1 - \omega) \sum_{i = 1}^{M} r_iZ_i,
\end{equation}  

\noindent subject to the constraint,

\begin{equation}
	\constraintFunction{asset}{\solution} = \sum_{i = 1}^{M} z_i.
\end{equation}

\noindent In this formulation, the Pauli-Z gates $Z_i$ encode a particular portfolio where, for each asset, eigenvalue $z_i \in \set{1,-1,0}$ represents a choice of long, short or no position. Associated with each asset is the expected return $r_i$ and covariance $\sigma_{ij}$ between assets $i$ and $j$; which are calculated using historical data. The risk parameter, $\omega$, weights consideration of $r_i$ and $\sigma_{ij}$ such that as $\omega \rightarrow 0$ the optimal portfolio is one providing maximum returns. In contrast, as $\omega \rightarrow 1$, the optimal portfolio is the one that minimises risk. The constraint $\constraintFunction{asset}{\solution}$ works to maintain the relative net position with respect to a pre-existing portfolio \cite{slate_quantum_2021}.

In the following examples, we demonstrate the application of the \qwoa{} and \qaoaz{} to a small `portfolio' consisting of four assets taken from the ASX 100, under the constraint $\constraintFunction{asset}{\solution} = 2$.

\subsubsection{Comparison of the \qwoa{} and \qaoaz{}.}

As outlined in \cref{sec:qwoa}, QWOA uses an indexing unitary $\Uindex$ to encode constraints on the solution space. As \quop{} is interested only in the unitary dynamics of the quantum state evolution, implementation of the indexing unitary simply requires that the user supply a quality function that returns the restricted solution space in a consistent order. For the \qwoa{} example, the set of valid solutions was calculated using the module `qwoa\_qualities.py', which was originally written for \cite{slate_quantum_2021}. It makes use of the pandas-webreader package \cite{noauthor_pandas-datareader_nodate} to source the daily adjusted close price of a given list of stocks from the Yahoo Finance website \cite{noauthor_yahoo_nodate}. When run as the main program, `qwoa\_qualities.py' returns the $\qualityVector$ corresponding to solutions in $\validSolutionSpace{}$ using historical data between user-specified dates and outputs $\qualityVector{}$ to a CSV file. For this example, the stocks AMP.AX, ANZ.AX and AMC.AX were considered between 1/1/2017 and 12/31/2018.

The \qwoa{} algorithm is included with \quop{} as a predefined module. As such, its simulation, shown in Example \ref{ex-4}, is similar in structure to the \qaoa{} program outlined in \cref{sec:example_qaoa}. However, as in this instance, because $\qualityVector{}$ is stored in a CSV file, we use the external package Pandas to read the quality values and the \verb|diagonal| submodule \verb|operator| function \verb|serial_array| to pass these values to the \verb|set_qualities| method on Lines 12 to 14. 

 Note that the size of the simulation is determined by the number of valid solutions $\size{\validSolutionSpace{}}$. This is distinct from a quantum implementation of the \qwoa{} algorithm as, while its $\U{mix}$ occurs over $\validSolutionSpace{}$, its $\U{phase}$ still acts on $\mathcal{S}$. However, because $\initialState{QWOA}$ is initialised as an equal superposition over $\validSolutionSpace{}$, quantum states associated with invalid solutions do not influence the idealised quantum dynamics. Hence, we can gain a significant performance advantage at no cost to simulation accuracy by restricting the classical simulation to $\validSolutionSpace{}$. 

In \cref{sec:example_qaoa}, the final state $\ansatz{}$ and $\qualityVector$ were saved to an HDF5 file and analysis of algorithm performance was determined via computation on these arrays. Studying the complete quantum state is essential to understanding the dynamics associated with a particular QVA application. Still, often a researcher is concerned more immediately with $\finalValue{}$ with respect to changes in $D$ or $\size{\validSolutionSpace{}}$. For this reason \quop{} supports the recording of important simulation metrics in a log file. Created via the \verb|set_log| method on Lines 16 to 19, the first argument specifies the name of the CSV output log file, the second argument specifies the simulation label, and the third argument specifies the write action, which follows the convention of \verb|a| to append or \verb|w| to (over)write. With a log file set the system size $N$, ansatz depth $D$, optimised objective function value $\finalValue{}$, state norm $\braket{\variationalParameters_f}_\text{}$, in-program simulation time, MPI communicator size, number of $\ansatz{}$ evaluations and the success status of the optimiser are recorded for each simulation instance.

To study how $\finalValue{\qwoa{}}$ changes as $D$ increases we call the \verb|benchmark| method on Lines 21 to 26. The first argument is an iterable object that provides a sequence of $D$ values, the second is the number of repeat simulations at each $D$. The keyword arguments \verb|filename| and \verb|save_action| specify that $\ansatzFinal{\qwoa{}}$ and $\qualityVector{}$ be saved to the new HDF5 file `qwoa\_portfolio.h5'. The \verb|param_persist| argument specifies a schema for $\variationalParameters_0$. If \verb|True|, for $D>1$ the best-performing $\variationalParameters$ at depth $D$ are used as the $\variationalParameters_0$ for the first $D$ ansatz iterations at depth $D + 1$.

A \qaoaz{} approach to the portfolio optimisation problem uses two parity mixers that act on the short and long qubits, respectively, such that the $\solutionSpace$ is partitioned into subgraphs of the same $\constraintFunction{asset}{\solution}$ value. For this example, we are considering four assets so the two parity mixers act on separate subspaces of $\mathcal{H}$ as shown below:
\hfill
\begin{center}
\begin{tikzpicture}[long/.style = {draw, circle}, short/.style = {draw, circle, dashed}, node distance= 0.15cm and 0cm]

	\node[long] (0,0) (one) {$\ket{l}$};
		\node[short] [right = of one] (two) {$\ket{s}$};
	\node[long] [right = of two] (three) {$\ket{l}$};
		\node[short] [right = of three] (four) {$\ket{s}$};
	\node[long] [right = of four] (five) {$\ket{l}$};
		\node[short] [right = of five] (six) {$\ket{s}$};
	\node[long] [right = of six] (seven) {$\ket{l}$};
		\node[short] [right = of seven] (eight) {$\ket{s}$};
		
	\node [below = of one] (0) {0};
	\node [below = of two] (1) {1};
	\node [below = of three] (2) {2};
	\node [below = of four] (3) {3};
	\node [below = of five] (4) {4};
	\node [below = of six] (5) {5};
	\node [below = of seven] (6) {6};
	\node [below = of eight] (7) {7};

\end{tikzpicture}
\end{center}
\hfill
\noindent Where $\ket{l}$ denotes a `long' qubit, $\ket{s}$ denotes a `short' qubit, and the numbering indicates the global index of each qubit.

To constrain probability amplitude to $\validSolutionSpace{}$, $\initialState{\qaoaz{}}$ is prepared as

\begin{equation}
	\initialState{\qaoaz{}} = \ket{01}^{\otimes A}\left( \frac{1}{\sqrt{2}}(\ket{00} + \ket{11})^{2N-A} \right),
\end{equation}

\noindent where $A$ is the desired value of $\constraintFunction{asset}{\solution}$. This creates a (non-equal) superposition of states across all qubit subgraphs with a net parity of $A$.

To implement this algorithm in \quop{} we use the \verb|Ansatz| class, the \verb|sparse| \verb|propagator| submodule, the \verb|diagonal| \verb|propagator| submodule, the \verb|observable| submodule, \verb|state| submodule and \verb|param| submodule. The \verb|toolkit| functions \verb|string|, \verb|X| and \verb|Y| are also used to define the parity mixers and $\initialState{\qaoaz{}}$.

As with Example \ref{ex-4}, the quality function is located in the external module `qaoaz\_portfolio.py', which is included in \quop{}/examples/portfolio. It follows the same method as `qwoz\_portflio.py', differing in that it has been written as a parallel quality function (see \cref{sec:user_fun}) that returns local partitions of the complete $\solutionSpace$.

The dual parity mixing operators are defined over three functions. The first of these (Lines 8 to 11) defines a generalisation of the Pauli-matrix terms used for the $B_\text{odd}$, $B_\text{even}$ and $B_\text{last}$ mixing operations in \cref{eq:mixing_qaoa}. The second function (Lines 13 to 41) takes a list of qubit indexes specifying a subspace of $\mathcal{H}$ and the total number of qubits as its arguments and returns $B_\text{odd}$, $B_\text{even}$ and $B_\text{last}$ acting on the subspace. The third function, \verb|portfolio_mixer| (Lines 43 to 48), takes a number of qubits as its argument. It partitions the input number of qubits into subgroups, as depicted in  \cref{fig:qaoaz_mixer}, and returns a list containing the six mixing operators in the SciPy CSR sparse matrix format. 

A function to generate $\initialState{portfolio}$ is defined on lines 50 to 57 where, on Lines 52 and 53, the \verb|kron_power| function takes NumPy array $a$ and integer $N$ as its arguments and returns $a^{\otimes N}$ and, on Line 54, the \verb|string| function generates a qubit state from its bit-string representation. The function \verb|kron|, on Line 56, takes a list of arrays and returns their tensor product.

We then proceed to the definition of $\hat{U}_\text{QAOAz}$ using the \verb|Ansatz| class. An initial state other than an equal superposition is specified using the \verb|set_initial_state| method on Lines 77 to 80. It follows the same input convention as the previously described `set' methods. The wrapper function \verb|state.serial| is used to parse and distribute the output of the serial \verb|parity_state| function.

As, in this instance, the diagonal of the phase-shift matrix exponent is equal to $\qualityVector{}$, the objective function is defined by calling \verb|set_observables| on Line 82 with an integer argument that specifies the position of the mixing operator in the input list of unitaries (Line 75).

Finally, an output log is specified, and the benchmark method is called to trial the QVA over the same range of $D$ and number of repeats as  the \qwoa{} simulations. The \verb|benchmark| method generates a reproducible sequence of integers used as random seeds for all \verb|param| functions in the \verb|param| submodule. In this way, we ensure that the \qwoa{} and \qaoaz{} simulations are carried out over the same set of $\variationalParameters_0$ at the starting ansatz depth of $D = 1$. 

\begin{figure}[H]
    \centering
	\footnotesize{\quop{}/examples/portfolio\_rebalancing/portfolio\_plots.py}
	\includegraphics[width=0.75\columnwidth]{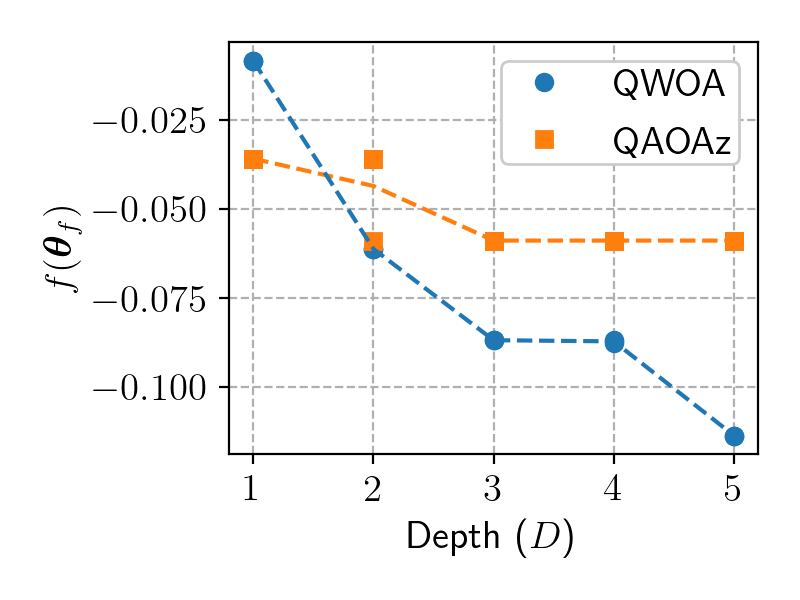}
	\caption{Optimised objective function value $\finalValue{}$ for the portfolio rebalancing problem using the \qwoa{} and \qaoaz{}.}
	\label{fig:portfolio}
\end{figure}

A comparison of the two algorithms is shown in \cref{fig:portfolio} where the $\finalValue{}$ was taken from the log file. For this brief comparison the \qwoa{} $\finalValue{}$ outperforms the \qaoaz{} for all $D > 1$.

\begin{figure*}[t]

	\begin{subfigure}[t]{0.245\textwidth}
	    \centering
		\includegraphics[width=\textwidth]{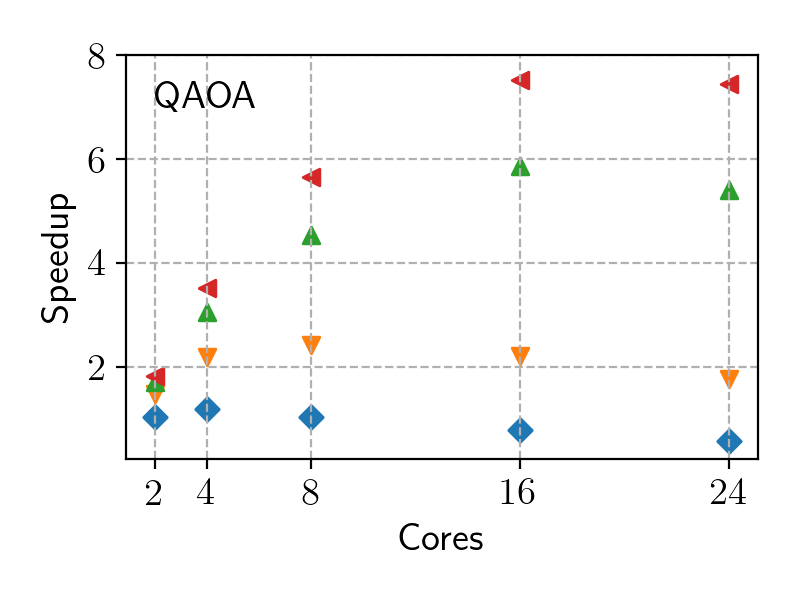}
	\end{subfigure}
	\hfill
	\begin{subfigure}[t]{0.245\textwidth}
		\centering
		\includegraphics[width=\textwidth]{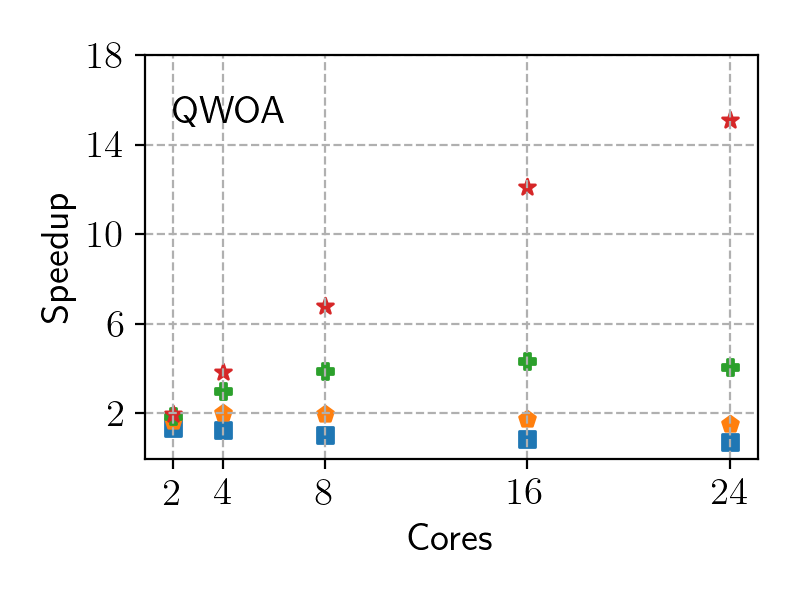}
	\end{subfigure}
	\hfill
	\begin{subfigure}[t]{0.245\textwidth}
		\centering
		\includegraphics[width=\textwidth]{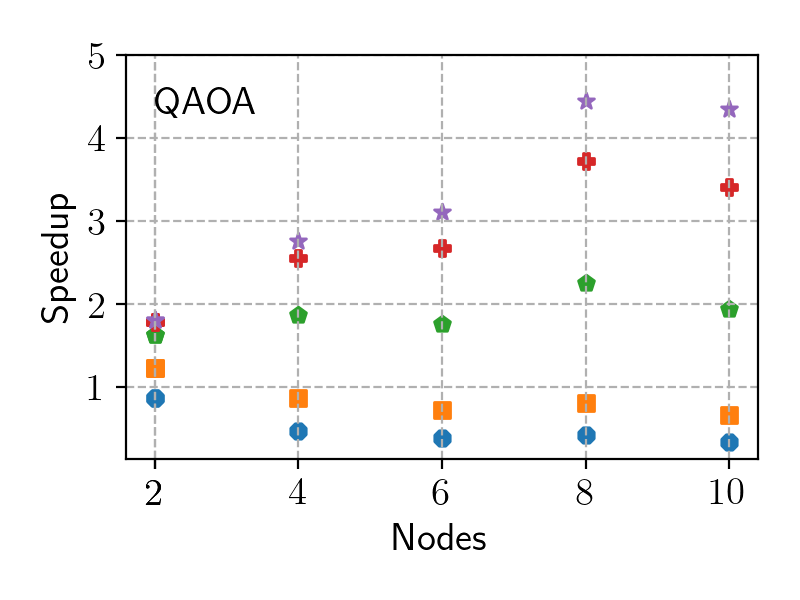}
	\end{subfigure}
	\hfill
	\begin{subfigure}[t]{0.245\textwidth}
		\centering
		\includegraphics[width=\textwidth]{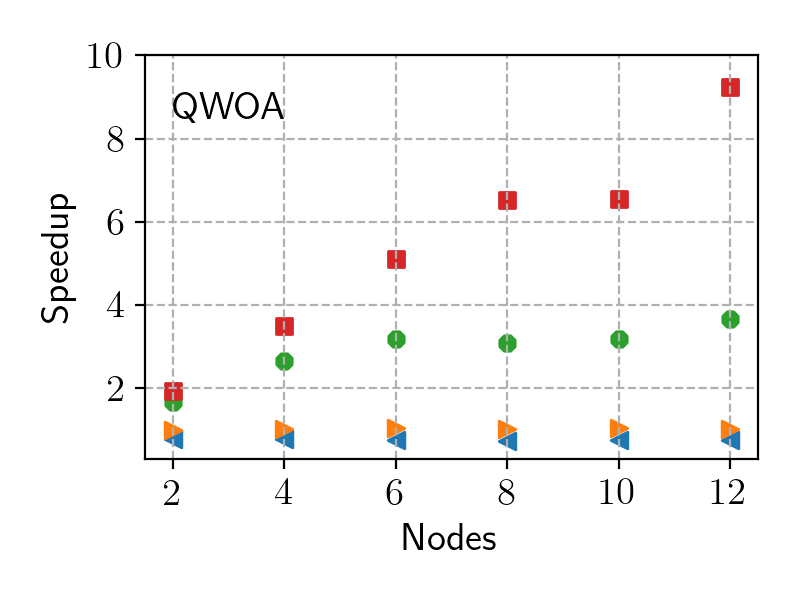}
	\end{subfigure}
	
	\begin{subfigure}[t]{0.245\textwidth}
	    \centering
		\includegraphics[width=\textwidth]{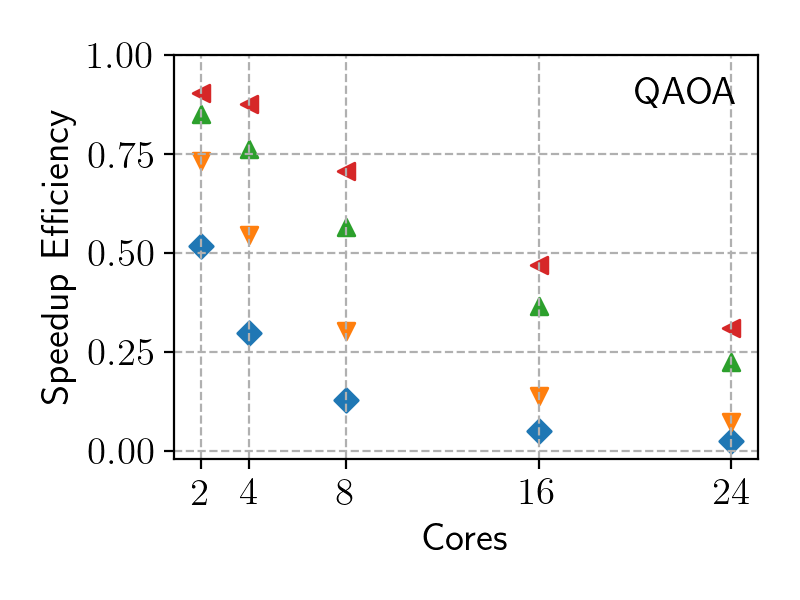}
	\end{subfigure}
	\hfill
	\begin{subfigure}[t]{0.245\textwidth}
		\centering
		\includegraphics[width=\textwidth]{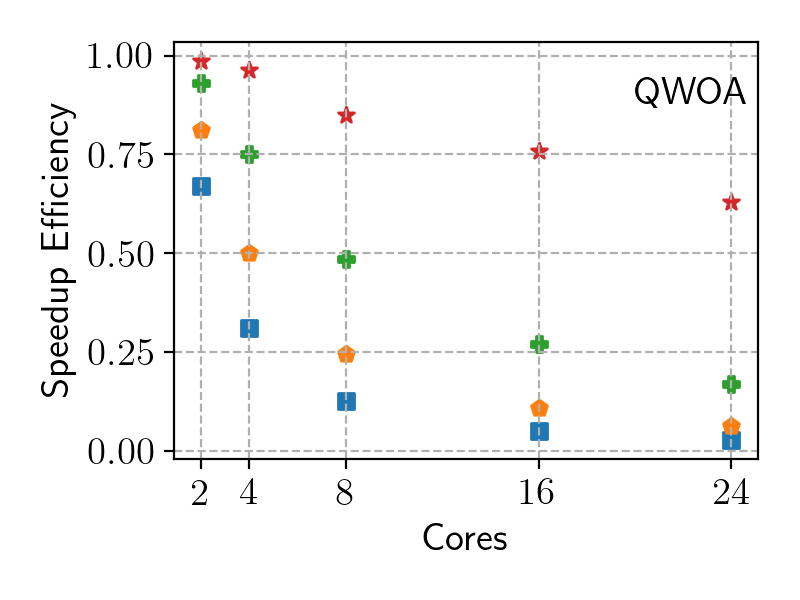}
	\end{subfigure}
	\hfill
	\begin{subfigure}[t]{0.245\textwidth}
		\centering
		\includegraphics[width=\textwidth]{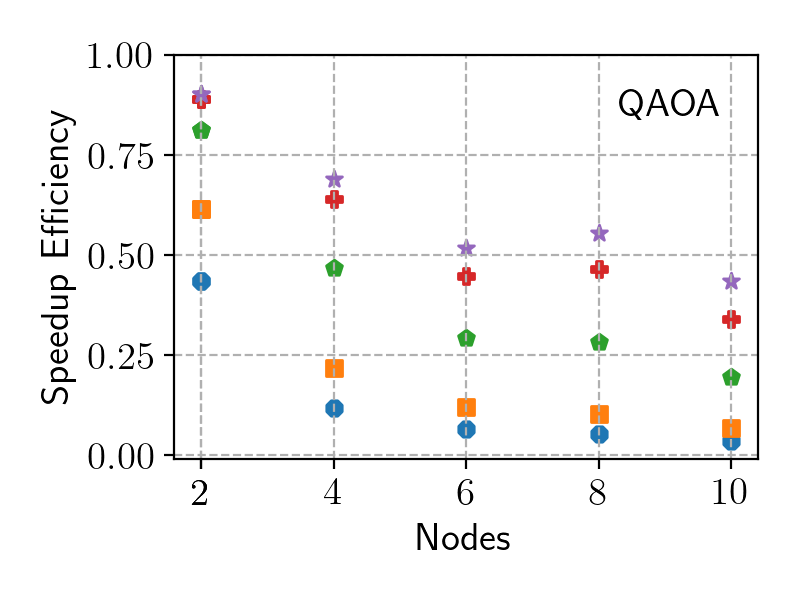}
	\end{subfigure}
	\hfill
	\begin{subfigure}[t]{0.245\textwidth}
		\centering
		\includegraphics[width=\textwidth]{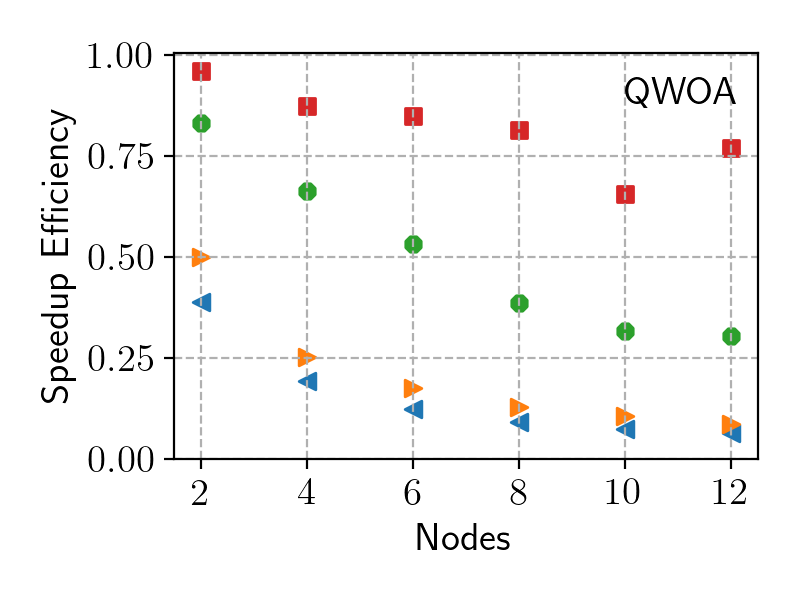}
	\end{subfigure}
    
	\begin{subfigure}[t]{0.245\textwidth}
	    \centering
		\includegraphics[width=0.8\textwidth]{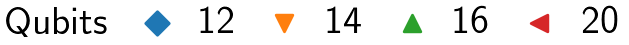}
		\caption{\label{fig:qaoa_workstation_strong}}
	\end{subfigure}
	\hfill
		\begin{subfigure}[t]{0.245\textwidth}
	    \centering
		\includegraphics[width=0.8\textwidth]{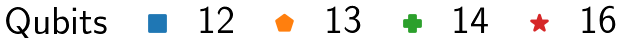}
		\caption{		\label{fig:qwoa_workstation_strong}}
	\end{subfigure}
	\hfill
	\begin{subfigure}[t]{0.245\textwidth}
	    \centering
		\includegraphics[width=0.95\textwidth]{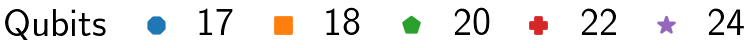}
		\caption{\label{fig:qaoa_cluster_strong}}
	\end{subfigure}
	\hfill
	\begin{subfigure}[t]{0.245\textwidth}
	    \centering
		\includegraphics[width=0.8\textwidth]{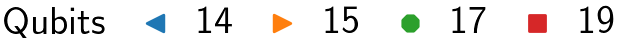}
		\caption{\label{fig:qwoa_cluster_strong}}
	\end{subfigure}

	\cprotect\caption{Strong scaling speedup and efficiency for the \verb|qaoa| and \verb|qwoa| state evolution methods running on a single and multiple nodes. For each trial, a \verb|qaoa| or \verb|qwoa| instance was instantiated with a $\qualityVector{}$ consisting of uniformly distributed floats in $(0,1]$. The ansatz depth was set to $D = 15$ such that calling the \verb|Ansatz| \verb|evolve_state| method resulted in 15 repeats of the state evolution subroutines implementing the phase-shift and mixing-unitaries. The $\variationalParameters_0$ were prepared identically for all trials at the same number of qubits from the uniform distribution $(0, 2 \pi]$. For (a) and (b) speedup is reported proportional to the time taken using a single CPU core (1 MPI process). The \verb|qaoa| the single-node wall-times were 3.27 s, 5.35 s, 17.3 s and 265 s for 12, 13, 14 and 16 qubits respectively and, for \verb|qwoa|, 2.39 s, 2.50 s, 5.05 s, 24.0 s and 727 s for 10, 12, 14, 16 and 20 qubits respectively. Efficiency is defined as the speedup divided by the number of CPU cores. For (c) and (d) all trails run on fully occupied nodes and the reported speedup is proportional to the time taken on one fully occupied node (24 MPI processes). For \verb|qaoa| the single-node wall-times were 7.64 s, 17.8 s, 97.5 s, 496 s and 2365 s for 12, 13, 14 and 16 qubits respectively and, for \verb|qwoa|, 4.26 s, 7.03 s, 60.0 s and 967 s for 10, 12, 14, 16 and 20 qubits respectively. Efficiency in (c) and (d) is defined as the speedup divided by the number of nodes.}
	\label{fig:strong}
\end{figure*}

\begin{figure*}[h!]
	\begin{subfigure}[t]{0.245\textwidth}
	    \centering
		\includegraphics[width=\textwidth]{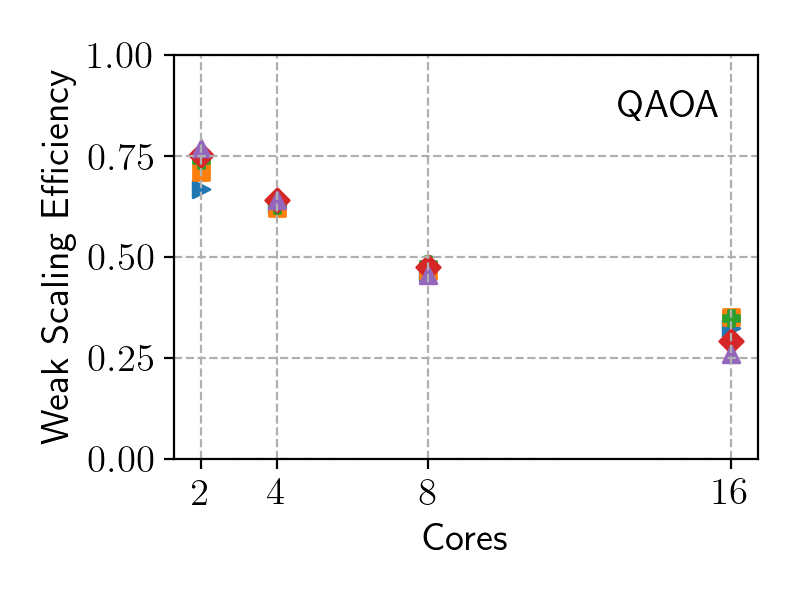}
	\end{subfigure}
	\hfill
	\begin{subfigure}[t]{0.245\textwidth}
		\centering
		\includegraphics[width=\textwidth]{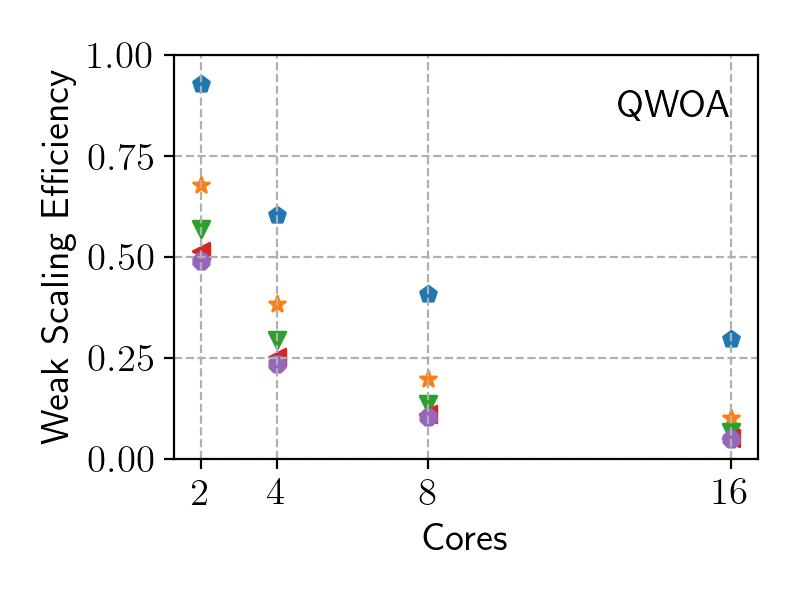}
	\end{subfigure}
	\hfill
	\begin{subfigure}[t]{0.245\textwidth}
	    \centering
		\includegraphics[width=\textwidth]{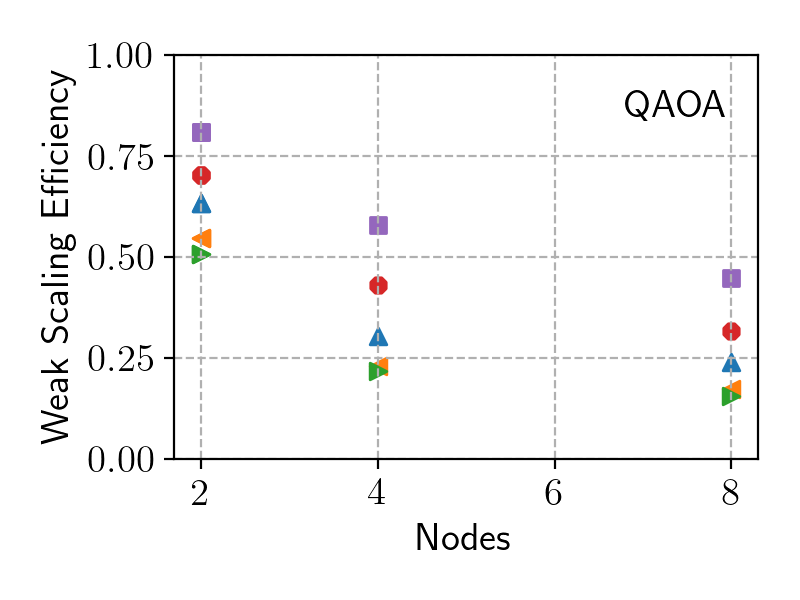}
	\end{subfigure}
	\hfill
	\begin{subfigure}[t]{0.245\textwidth}
		\centering
		\includegraphics[width=\textwidth]{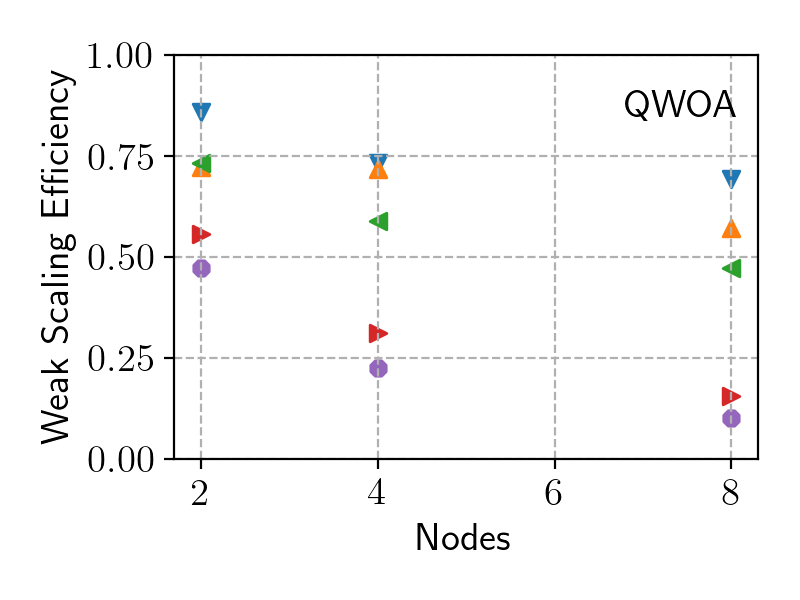}
	\end{subfigure}

	\begin{subfigure}[t]{0.245\textwidth}
	    \centering
		\includegraphics[width=0.85\textwidth]{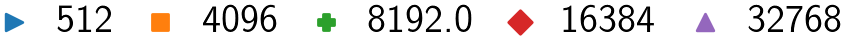}
		\caption{\label{fig:qaoa_weak_workstation}}
	\end{subfigure}
	\hfill
	\begin{subfigure}[t]{0.245\textwidth}
	    \centering
		\includegraphics[width=0.85\textwidth]{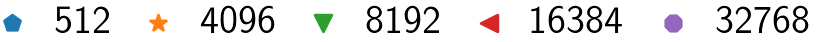}
		\caption{\label{fig:qwoa_weak_workstation}}
	\end{subfigure}
	\hfill
	\begin{subfigure}[t]{0.245\textwidth}
	    \centering
		\includegraphics[width=0.95\textwidth]{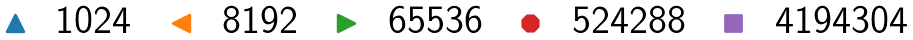}
		\caption{\label{fig:qaoa_weak_cluster}}
	\end{subfigure}
	\begin{subfigure}[t]{0.245\textwidth}
	    \centering
		\includegraphics[width=0.85\textwidth]{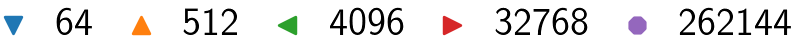}
		\caption{\label{fig:qwoa_weak_cluster}}
	\end{subfigure}
	\cprotect\caption{The weak scaling efficiency for the \verb|qaoa| and \verb|qwoa| state evolution methods with state vector partitions of size \verb|local_i| as indicated by the corresponding plot legends. For (a) and (b), efficiency is defined as  $T(1)/T(\text{Cores})$ where $T(1)$ is the wall-time for one MPI process with a system size of \verb|local_i|. For (c) and (d), efficiency is defined as $T(1)/T(\text{Nodes})$ where $T(1)$ is the wall-time for one Node of 24 MPI processes with \verb|local_i| state vector elements and all nodes were fully occupied at 24 MPI process per-node.}
	\label{fig:weak}
\end{figure*}

\begin{figure}
    \centering
	\begin{subfigure}[t]{0.49\columnwidth}
	    \centering
		\includegraphics[width=\textwidth]{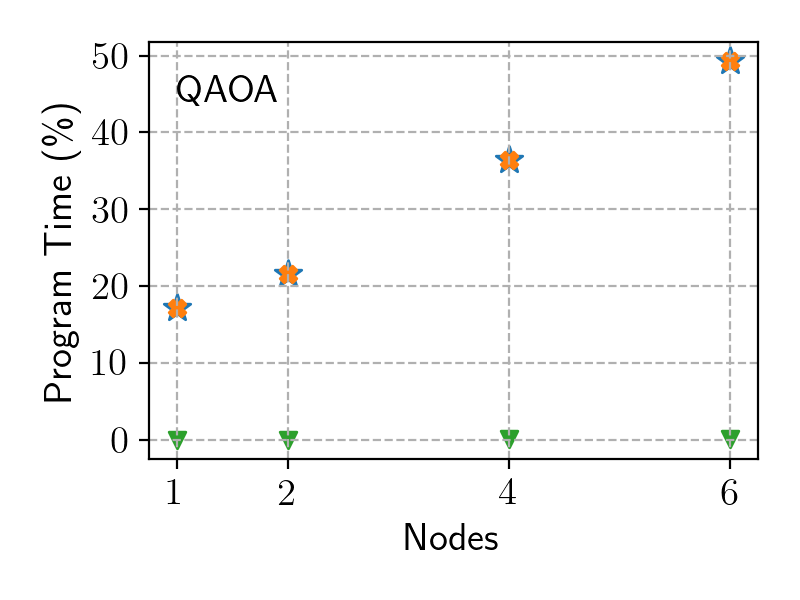}
	\end{subfigure}
	\begin{subfigure}[t]{0.49\columnwidth}
		\centering
		\includegraphics[width=\textwidth]{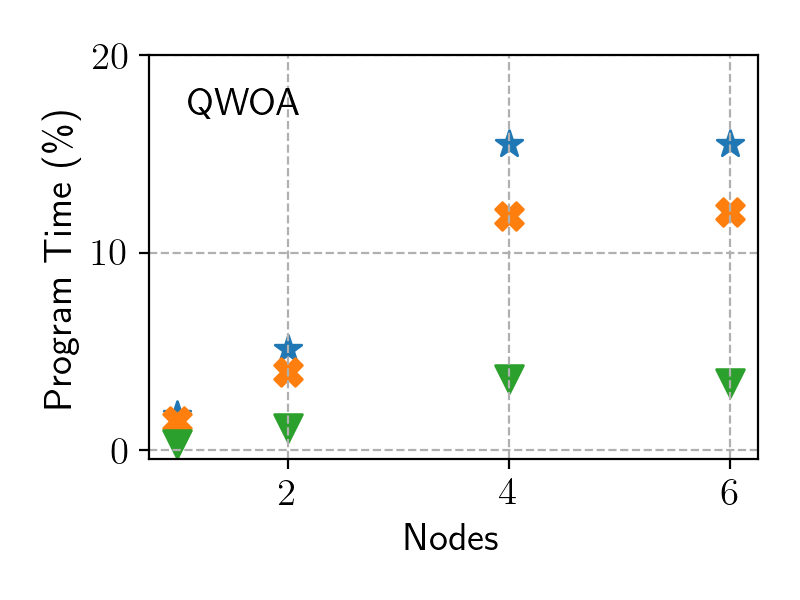}
	\end{subfigure}

	\begin{subfigure}[t]{\columnwidth}
	    \centering
		\includegraphics[width=0.65\textwidth]{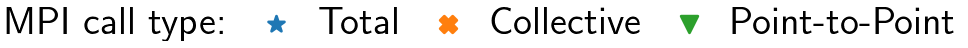}
	\end{subfigure}

	\cprotect\caption{The percentage of program wall-time spent in MPI calls for the \verb|qaoa| and \verb|qwoa| state evolution methods at 22 and 19 qubits, respectively, as reported by Arm Map version 19.0.1. All nodes were fully occupied at 24 MPI processes per node.}
	\label{fig:evolve_state_MPI_overhead}
\end{figure}

\begin{figure}[]

	\begin{subfigure}{0.495\columnwidth}
	    \centering
		\includegraphics[width=\textwidth]{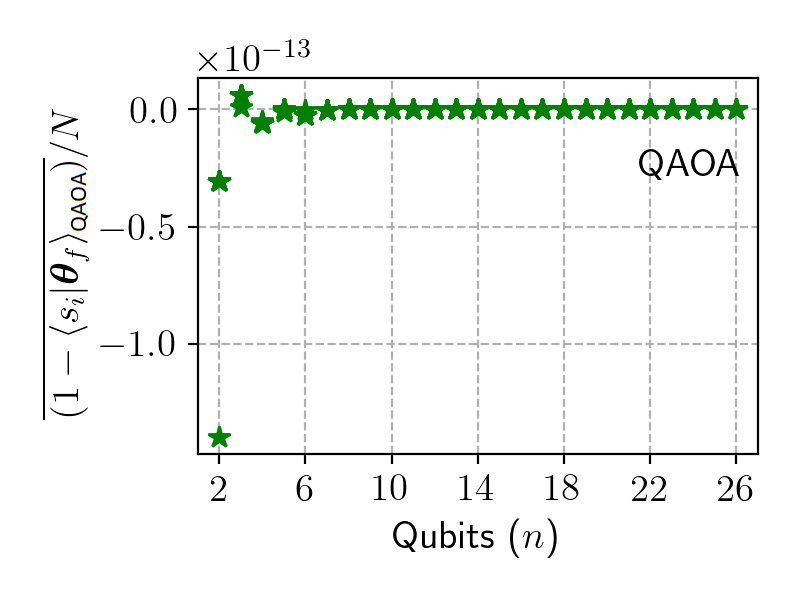}
	\end{subfigure}
	\hfill
	\begin{subfigure}{0.495\columnwidth}
	    \centering
		\includegraphics[width=\textwidth]{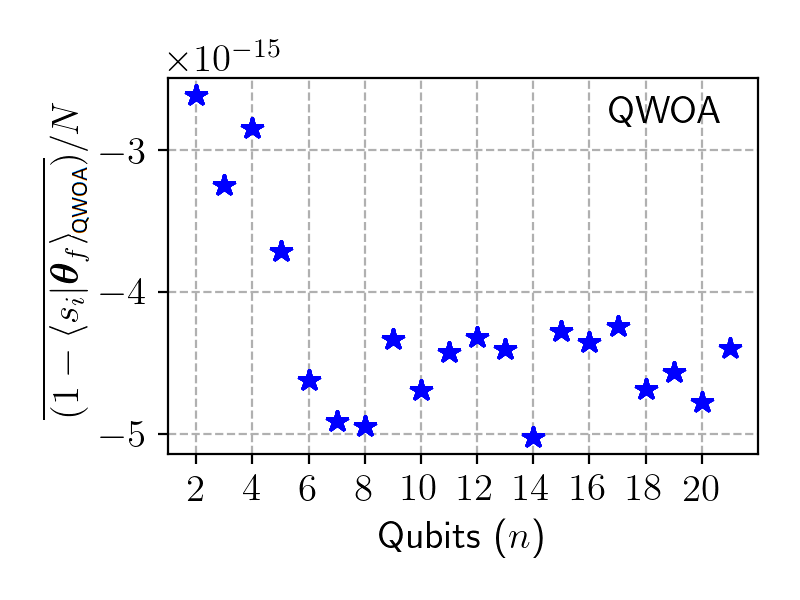}
	\end{subfigure}
	
	\caption{Average deviation from the norm for the $\ansatzFinal{QAOA}$ (left) and $\ansatzFinal{QWOA}$ (right) depicted in \cref{fig:strong,fig:weak}}
	\label{fig:norm_deviation}
	\vspace{2em}
\end{figure}

\section{Performance} \label{sec:performence}

The performance of \quop{} was assessed on the `Magnus' system at the Pawsey Supercomputing Centre - a Cray XC40 Series Supercomputer with an Aries interconnect rated at 72 gigabits-per-second per node. Each node consisted of two Intel Xeon E5-2690V3 `Haswell' CPUs with 12 cores clocked at 2.6 GHz and 64 GB of RAM.

The strong scaling behaviour of the \qaoa{} and \qwoa{} evolution methods on a single compute node is shown in  \cref{fig:qaoa_workstation_strong,fig:qwoa_workstation_strong}. As defined in \cref{eq:qaoa,eq:qwoa}, each of the unitaries contains a phase-shift-unitary followed by a mixing-unitary. For $\ansatz{\qaoa{}}$ these are implemented using the \verb|diagonal| and \verb|sparse| unitary classes and, for $\ansatz{\qwoa{}}$,  the \verb|diagonal| and \verb|circulant| unitary classes. A parallel advantage in \qaoa{} state evolution is observed for systems of at least 12 qubits, with a system of 20 qubits scaling with an efficiency greater than 0.5 up to eight CPU cores. For \qwoa{}, parallel advantage beyond two CPU cores starts at 12 qubits, with a system of 16 qubits achieving a speedup of 15.1 with an efficiency of 0.63 at 24 CPU cores. 

Strong scaling behaviour for \qaoa{} and \qwoa{} evolution across multiple nodes is shown in \cref{fig:qaoa_cluster_strong,fig:qwoa_cluster_strong}. Efficient scaling of state evolution to two nodes occurs at 17 qubits for \qaoa{} and 15 qubits for \qwoa{}. For \qaoa{} at 24 qubits, a speedup of 5.23 times at an efficiency of 0.44 was achieved at 12 nodes (288  cores) with respect to the wall-time of a fully occupied single node (24 cores). For \qwoa{}, the equivalent comparison shows a speedup of 9.25 with an efficiency of 0.77 at 12 nodes.

The \qwoa{} and \qaoa{} state evolution methods were profiled using Arm Map 19.0.1  to quantify the degree of communication overhead in a distributed computing environment as shown by \cref{fig:evolve_state_MPI_overhead}.  For \qaoa{} state evolution at 22 qubits, the overhead ranged from a total of 17.2 \% at one node and 49.3 \% at six nodes. The increase in communication overhead is responsible for the decrease in strong scaling efficiency with \qaoa{} state evolution at 22 qubits having an efficiency that falls below 0.5 for nodes greater than six (see \cref{fig:qaoa_cluster_strong}). Almost all of the \qaoa{} state evolution MPI call time is spent in collective calls, of which the majority are `Alltoallv' calls responsible for the sending and receiving of state vector elements during matrix multiplication. State evolution for the \qwoa{} is dominated by a one-dimensional Fourier transform computed in MPI parallel using the FFTW3 package. For \qwoa{} state evolution at 19 qubits, the time spent in MPI calls ranges from 4.8 \% for one node and 19.6 \% at six nodes, a modest increase that is in line with the  efficient scaling depicted in \cref{fig:qwoa_cluster_strong}. 

Scalability of the state evolution methods for \verb|qaoa| and \verb|qwoa| at a constant MPI process load (weak scaling) is shown in \cref{fig:weak} with respect to cores on a single node and a cluster of multiple nodes. In each instance, imperfect weak scaling is observed as, for both \qaoa{} and \qwoa{}, increases in $n$ are accompanied by an increased degree of inter-qubit coupling. For high process loads, the \verb|qaoa| state evolution method scales more efficiently than \verb|qwoa| state evolution, which is consistent with the structure of the corresponding mixing operators. For the \qaoa{} the hypercube matrix operator has a sparse banded-diagonal structure that, for $\text{local}_i \mod 2 = 0$, induces a communication overhead of $\mathcal{O}(\log_2(\mpiSize))$. In contrast, the complete-graph mixing operator of the \qwoa{} requires communication between all of the MPI processes resulting a communication overhead of $\mathcal{O}(\mpiSize)$. 

As a measure of solution quality, deviation from the norm was calculated for the state evolution results shown in \cref{fig:strong,fig:weak}. Figure \ref{fig:norm_deviation} shows the total deviation divided by the system size to indicate the per-state accuracy. For both \qaoa{} and \qwoa{} the deviation is on the order of $10^{-13}$ or below, which is consistent with double precision accuracy.

\begin{figure}[t!]
    \centering
	\begin{subfigure}{\columnwidth}
    \centering
		\includegraphics[width=0.75\columnwidth]{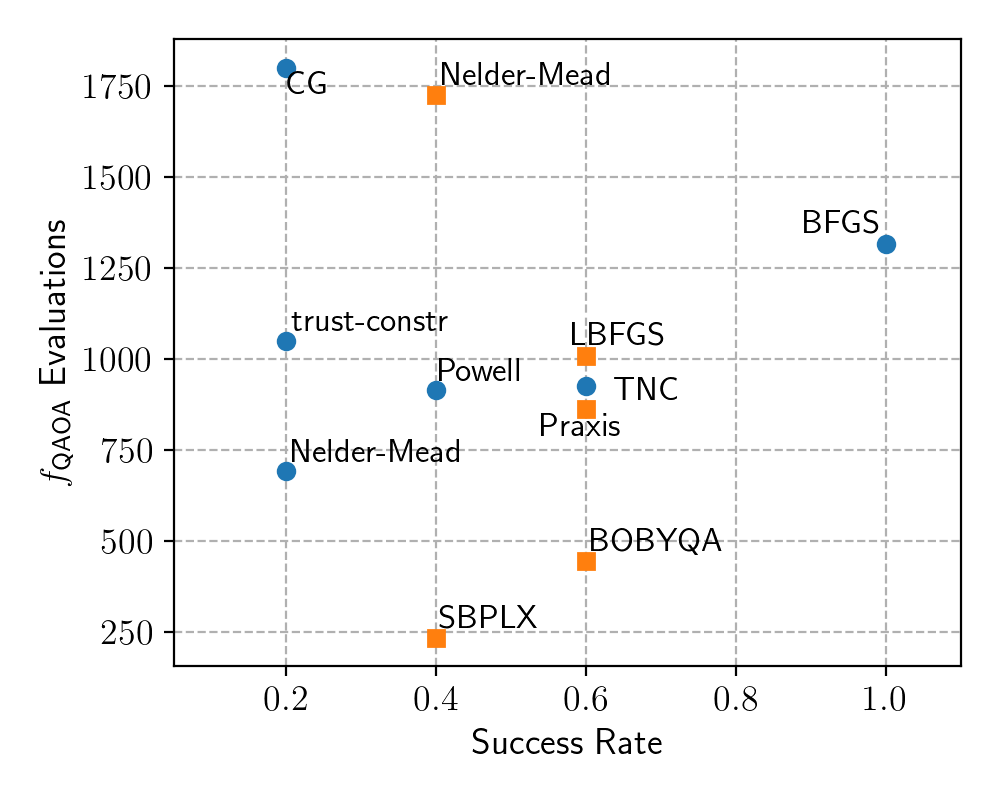}
	\end{subfigure}
	
	\begin{subfigure}{\columnwidth}
    \centering
		\includegraphics[width=0.75\columnwidth]{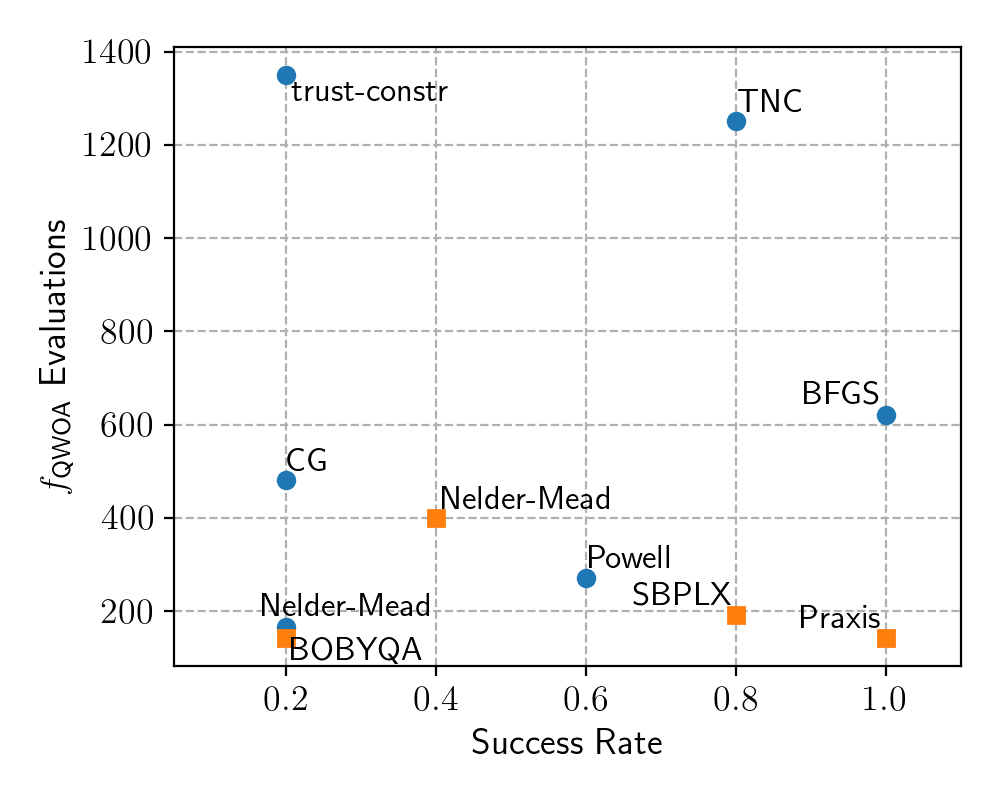}
	\end{subfigure}
	
	\begin{subfigure}{\columnwidth}
	\centering
		\includegraphics[width=0.4\columnwidth]{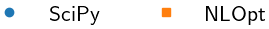}
	\end{subfigure}
	
	\caption{A comparison of the optimisation algorithms included with the SciPy and NLOpt packages. The plots depict algorithms that satisfied the convergence criteria in at least one out of the five trials. The complete list of considered algorithms is SciPy: BFGS, CG, Nelder-Mead, trust-constr, Powell, TNC and NLOpt: LD\_LBFGS, LN\_BOBYQA, LN\_PRAXIS, LN\_NELDERMEAD, LN\_SBPLX, LD\_MMA, LD\_CCSAQ. The comparison was carried out on 17 nodes over 48 hours.}
	\label{fig:optimiser_comparison}
\end{figure}

The effectiveness of the various optimisation algorithms included with the SciPy and NLOpt packages was considered with respect to simulation of the \qaoa{} and \qwoa{}. This comparison adopted methodology outlined in the NLOpt documentation  \cite{johnson_nlopt_nodate}. A system of 16 qubits ($N = 2^{16}$) was considered with a randomly generated  $\qualityVector{}$ consisting of values from a uniform distribution over $(0,1]$. Five sets of $\variationalParameters_0$ were generated for $D = 5$ ($\size{\variationalParameters{}} = 10$) and the algorithms simulated using the optimisers listed in the caption of \cref{fig:optimiser_comparison}. For each of the five sets of $\variationalParameters_0$ the lowest $\finalTheta{}$ was used to define five instances of the modified objective function

\begin{equation}
    f^\prime(\variationalParameters) = \size{\min(f_{\variationalParameters_i}) - \objectiveFunction{}},
    \label{eq:optimiser_objective}
\end{equation}

\noindent where $\min(f_{\variationalParameters_i})$ is the minimum $\finalValue{}$ found by any of the considered optimisation algorithms with initial variational parameters $\variationalParameters_i$. Each optimisation algorithm was then trialled with starting parameters $\variationalParameters_i$ and the objective function defined as in \cref{eq:optimiser_objective}. A particular algorithm was considered to have `succeeded' if it converged to a point satisfying $f^\prime(\variationalParameters) < b$, where $b = 0.8$ was chosen as it produced an informative measure across a large subset of the considered optimisers.

For the \qaoa{} trials the minimum final objective function values $\finalValue{QAOA}$ were 0.162 (Powell), 0.156 (BFGS), 0.120 (BOBYQA), 0.228 (LD\_LBFGS) and 0.154 (LD\_LBFGS). For the \qwoa{} the $\finalValue{QWOA}$ were  0.074 (BFGS), 0.086 (LN\_SBPLX), 0.078 (BFGS), 0.06 (BFGS) and 0.073 (Powell). As shown in \cref{fig:optimiser_comparison}, BFGS was the only algorithm which consistently satisfied the convergence test for both the \qaoa{} and the \qwoa{}. This result, in combination with a relatively low number of associated $\ansatz{}$ evaluations, supports the use of BFGS as the default \quop{} optimisation algorithm.

\begin{figure}[t!]
	 \centering
	\begin{subfigure}[t]{0.495\columnwidth}
	    \centering
		\includegraphics[width=\textwidth]{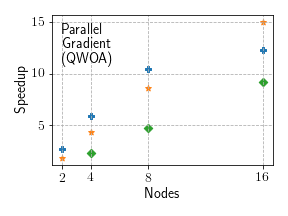}
	\end{subfigure}
    \hfill
	\begin{subfigure}[t]{0.495\columnwidth}
		\centering
		\includegraphics[width=\textwidth]{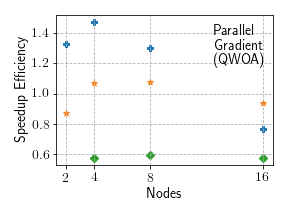}
	\end{subfigure}

	\begin{subfigure}[t]{0.5\columnwidth}
	    \centering
		\includegraphics[width=0.7\columnwidth]{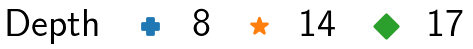}
	\end{subfigure}

	\caption{Speedup achieved with parallel computation of $\gradient$ for the \qwoa{} at 16 qubits with the $\qualityVector$ and $\variationalParameters_0$ defined as described in \cref{fig:strong}. Each node introduced an additional $\mpiAlg$ sub-communicator with 24 MPI processes.}
	\label{fig:parallel-gradient-scaling}
\end{figure}

\begin{figure}[t!]
	\centering
	\begin{subfigure}[t]{\columnwidth}
		\centering
		\includegraphics[width=0.6\columnwidth]{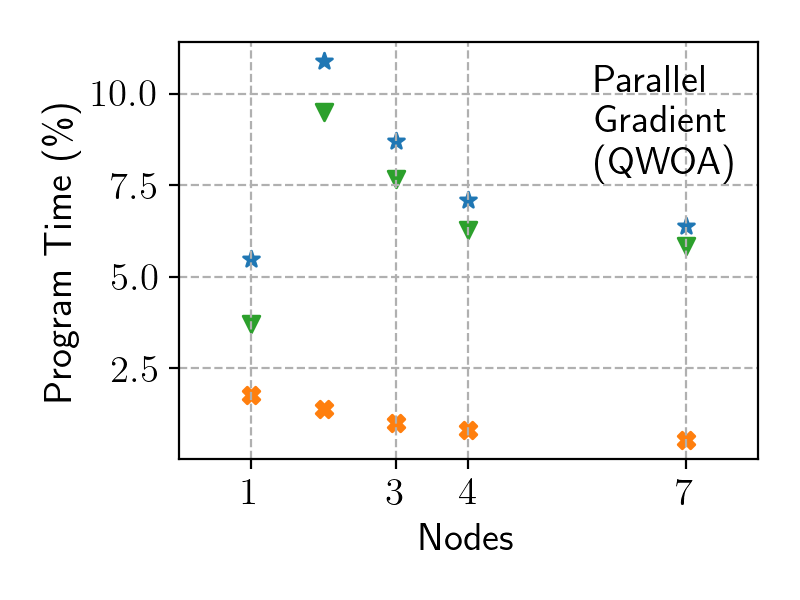}
	\end{subfigure}

	\begin{subfigure}[t]{\columnwidth}
	    \centering
		\includegraphics[width=0.8\columnwidth]{graphics/MPI_overhead_legend.png}
	\end{subfigure}

	\caption{The percentage of program wall-time spent in MPI calls for execution of the \qwoa{} at $D=14$ (see \cref{fig:parallel-gradient-scaling}) as reported by Arm Map version 19.0.1.}
	\label{fig:gradient_MPI_overhead}
\end{figure}

The scaling behaviour for parallel computation of the gradient $\gradient$ is shown in \cref{fig:parallel-gradient-scaling} for simulation of the \qwoa{} algorithm with 18 qubits at $D = 8, 14, 17$. This mode of parallelism scales very efficiently; at 16 nodes, there was a maximum speedup of 15.0 with an efficiency of 0.94 ($D = 14$) and a minimum speedup of 9.17 with an efficiency of 0.57 ($D = 8$). As shown by  \cref{fig:gradient_MPI_overhead}, additional nodes resulted in a negligible increase in MPI overhead as communication between sub-communicators consisted of only the updated $\variationalParameters$ and the partial derivatives of $\objectiveFunction{\qaoa{}}$. 

\begin{figure}
    \centering
	\includegraphics[width=0.75\columnwidth]{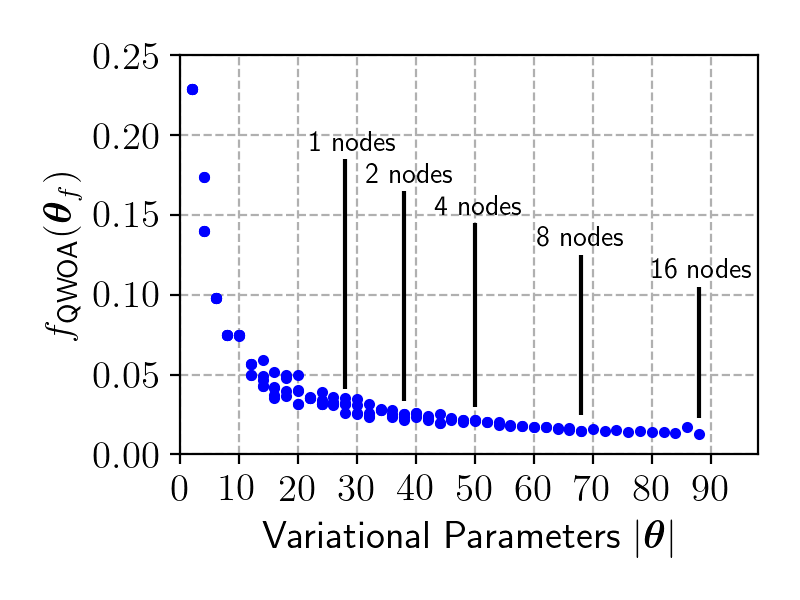}
	\caption{The optimised objective function $\finalValue{QWOA}$ for \qwoa{} simulations as described in \cref{fig:parallel-gradient-scaling} using 1, 2, 4, 8 and 16 compute nodes. Markers indicate the maximum number of $\variationalParameters$ simulated for the given number of nodes at a cumulative program wall-time of one hour.}
	\label{fig:optimisation-results}
\end{figure}

\clearpage

The convergence and simulation wall-time of \quop{} was compared to TensorFlow Quantum (TFQ); a \python{} package released in 2020 to support research in quantum-classical machine learning. This package was chosen for comparison as it targets a similar userbase through its approachable \python{} interface, focus on classically parameterised quantum algorithms and performant simulation of the complete wavefunction \cite{broughton_tensorflow_2020}. TensorFlow Quantum differs from \quop{} in two key areas. Firstly, it implements a gate-based approach to quantum simulation and, secondly, the package utilises a GPU accelerated library that computes $\ansatzFinal{QAOA}$ to single-precision accuracy \cite{tensor_github}.

The two packages were applied to simulation of the \qaoa{} applied to the max-cut problem for a regular random graph of degree three (see \cref{sec:usage}). Simulations were carried out at $D=2$ over two non-identical sets of five $\variationalParameters_0$ for all even $n$ in $[14,26]$. Optimisation was carried out for a maximum of 1000 $\objectiveFunction{QAOA}$ evaluations under the convergence criteria  $\Delta \objectiveFunction{QAOA} \leq 10^{-4}$. 

Implementation of \qaoa{} in TFQ built on an example included in the TFQ white-paper \cite{broughton_tensorflow_2020}. A quantum circuit exactly implementing \cref{eq:qaoa,eq:maxcut-cost} was generated using the \verb|tfq.util.exponential| function. This was used to define a Keras model with a single hidden layer for which $\initialState{QAOA}$ was passed to the input layer, and $\finalValue{QAOA}$ was returned by the output layer. The model was trained up to a maximum of 1000 epochs using the `Adam' optimiser, an absolute mean error loss function, and the training data-set $(\initialState{QAOA}, 0)$ (where $0$ is the minimum of \cref{eq:maxcut-cost}). The convergence criteria were implemented using an early-stopping callback function, stopping when the criteria were met over ten successive epochs.  

The \qaoa{} was implemented in \quop{} as shown in Example \ref{ex-1} with the $\qualityVector$ computed in parallel (see Example \ref{ex-3}). The  L-BFGS-B algorithm provided by SciPy was selected over the BFGS algorithm as it supported specification of the convergence criteria and maximum $\objectiveFunction{QAOA}$ evaluations via its \verb|ftol| and \verb|maxfun| options.     

Comparison simulations were carried out on a workstation (\quop{} and TFQ) and the `Magnus' cluster (\quop{} only). The workstation was equipped with an AMD Ryzen Threadripper 3970X 32-Core Processor at 3.7 GHz, 64 GB of RAM and an Nvidia RTX 3070 GPU. For all trials on the workstation, TFQ offloaded compute to the GPU and \quop{} ran with one MPI process per CPU core. Trials on the cluster ran on a variable number of nodes that were selected with reference to \cref{fig:strong}. 

\begin{figure}[t!]
	\centering
	\begin{subfigure}[t]{\columnwidth}
		\centering
		\includegraphics[width=0.98\columnwidth]{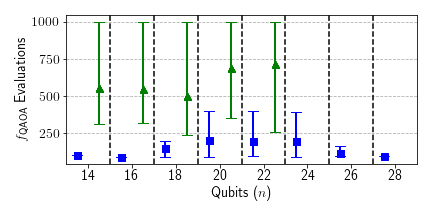}
		\caption{}
	\end{subfigure}
	
	\begin{subfigure}[t]{\columnwidth}
		\centering
		\includegraphics[width=0.98\columnwidth]{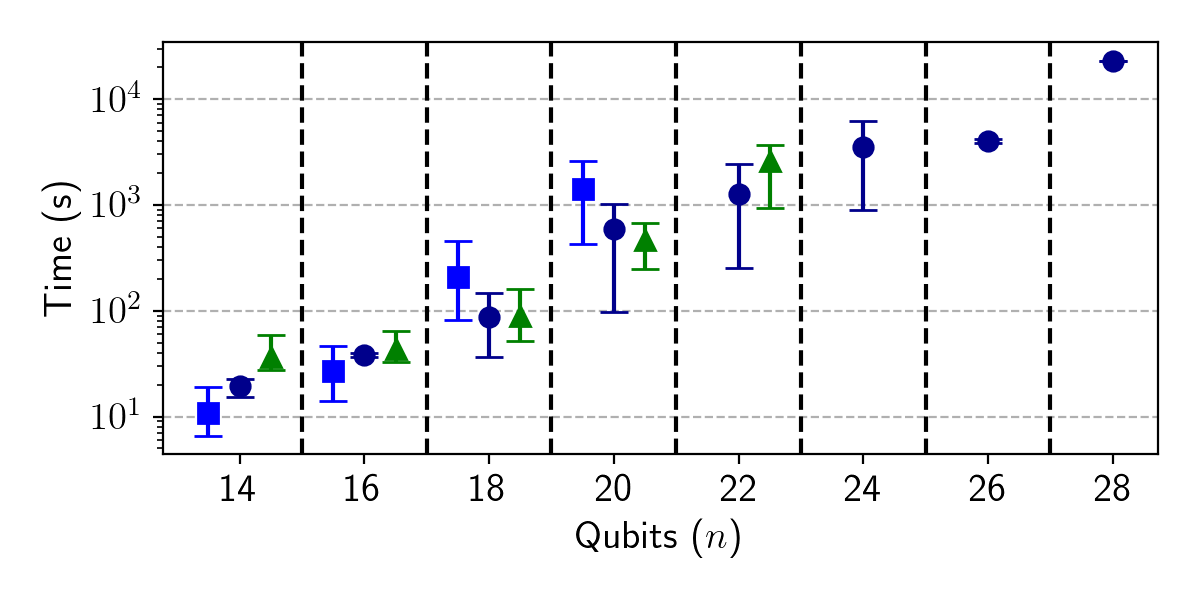}
	\end{subfigure}
	
	\begin{subfigure}[t]{\columnwidth}
	    \centering
		\includegraphics[width=0.8\columnwidth]{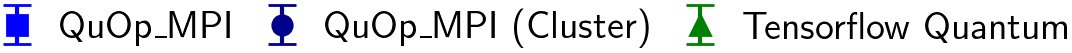}
		\caption{}
	\end{subfigure}

	\cprotect\caption{(a) Mean number of $\objectiveFunction{QAOA}$ evaluations and (b) mean program wall-time for simulation of the \qaoa{} as applied to the max-cut problem for regular random graphs of degree three using \quop{} on the workstation, \quop{} on the `Magnus' cluster and TensorFlow Quantum on the workstation. For (a) and (b) from 14 to 26 qubits the markers depict the mean value over sets of five $\variationalParameters_0$, the lower bar indicates the set minimum and the upper bar indicates the set maximum. The data shown for 28 qubits is for a single \quop{} simulation. On the workstation, simulations beyond 20 qubits for \quop{} and 22 qubits for TFQ were not possible due to memory constraints. The compute node configurations for \quop{} simulations on the cluster were 12 processes on one node for 14 qubits, 24 processes on one node at four qubits, 96 processes on two nodes for 18 qubits, 144 processes on six nodes for 20 qubits, 288 processes on 12 nodes for 22 qubits, 384 processes on 16 nodes for 24 qubits, 432 processes on 18 nodes for 26 qubits and 3360 processes on 140 nodes for 28 qubits.}
	\label{fig:quop_vs_tfq}
\end{figure}

Overall, the two packages performed similarly for minimisation of $\objectiveFunction{QAOA}$, with the lowest minima being 5.21 at 14 qubits (TFQ), 6.00 at 16 qubits (\quop{}), 6.80 at 18 qubits (\quop{}), 7.82 at 20 qubits (\quop{}) and 8.21 at 22 qubits (TFQ). \quop{} had an average $\finalValue{QAOA}$ of 5.56 at 14 qubits, 6.01 at 16 qubits, 7.00 at 18 qubits, 8.09 at 20 qubits and 8.59 at 22 qubits. The TFQ average $\finalValue{QAOA}$ were higher with 6.05 at 14 qubits, 6.70 at 16 qubits, 7.82 at 18 qubits, 8.63 at 20 qubits and 8.93 at 22 qubits. To investigate the source of this discrepancy, two sets of equivalent \quop{} max-cut simulations were carried out at 12, 14 and 16 qubits over sets of 50 $\variationalParameters_0$ with the $\objectiveFunction{QAOA}$ returned to single-precision for the first set and double-precision for the latter. Returning the objective function to double-precision accuracy (the \quop{} default) resulted in $\finalValue{QAOA}$ that were consistently lower than the $\finalValue{QAOA}$ obtained with a single-precision $\objectiveFunction{QAOA}$ (0.47 lower on average). This result indicates that the difference in simulation precision likely contributes to the observed difference in the mean $\finalValue{QAOA}$ between \quop{} and TFQ.

\cref{fig:quop_vs_tfq} depicts the mean number of $\objectiveFunction{QAOA}$ evaluations and the mean simulation wall-time for \quop{} and TFQ. Over the range of comparable simulations, \quop{} requires a smaller number of $\objectiveFunction{QAOA}$ evaluations. As such, \quop{} on the workstation has a simulation wall-time that is close to TFQ at 14 and 16 qubits. The simulation wall-time for TFQ at 18 and 20 qubits is significantly lower than \quop{} - which is consistent with TFQ's use of GPU acceleration and lower target precision. At 22 qubits, TFQ had an average wall-time of 2595 s, with the equivalent \quop{} simulation taking an average of 1267 s on 12 nodes (288 cores). This was the largest system simulated with TFQ, as simulations beyond this point were not possible due to GPU memory limitations. The distributed-memory parallelism of \quop{} allowed for simulations beyond 22 qubits with an average wall-time for 24 qubits of 3516 s on 16 compute nodes (384 cores) and, for 26 qubits, 4013 s on 18 nodes (432 cores). A single simulation at 28 qubits had a wall time of 23028 s on 140 compute nodes (3360 cores). Altogether these results demonstrate the utility of the high-precision simulation and scalable distributed memory parallelism of \quop{}. 

\section{Conclusion} \label{sec:conclusion}

\quop{} provides a highly scalable and flexible platform for parallel simulation and design of QVAs. As shown by example, researchers can quickly write programs to simulate several previously studied quantum optimisation algorithms, including the \qaoa{}, \extendedQaoa{}, \qwoa{} and \qaoaz{}, which are capable of running efficiently on massively parallel systems.

While this introduction to the package has focused on combinatorial optimisation following a pattern of alternating phase-shift and mixing-unitaries, the flexibility afforded of \quop{} allows for exploration of QVAs beyond this paradigm. Also not explored has been the application of \quop{} to the simulation of quantum variational eigensolver algorithms, which, while falling within the simulation framework of \quop{}, lie outside the immediate research interests of the authors.

Currently, \quop{} supports the efficient simulation of sparse and circulant mixing operators. While this covers the majority of mixing operators considered in the literature of QVAs, the scope of the package would be improved by the inclusion of a propagation method supporting dense mixing operators and a tensor network backend for the approximation of larger quantum systems. These features are slated for a future update. 

\section*{Acknowledgements}

This work was supported by resources provided by the Pawsey Supercomputing Centre with funding from the Australian Government and the Government of Western Australia. EM acknowledges the support of the Australian Government Research Training Program Scholarship. The authors would like to thank Sam Marsh, Nicholas Slate, Tavis Bennett, Mark Walker, Burbukje Shakjiri, Andrew Freedland, Zecheng Li, Yuhui Wang and Jianing Sun for their valuable feedback and code testing during the development of \quop{}.

\bibliographystyle{elsarticle-num}
\bibliography{quop}

\vspace{-3em}

\section*{}

\vspace{1em}

\setlength\intextsep{0pt}
\begin{wrapfigure}{l}{2.5cm}
\raggedright
\includegraphics[width=0.3\columnwidth]{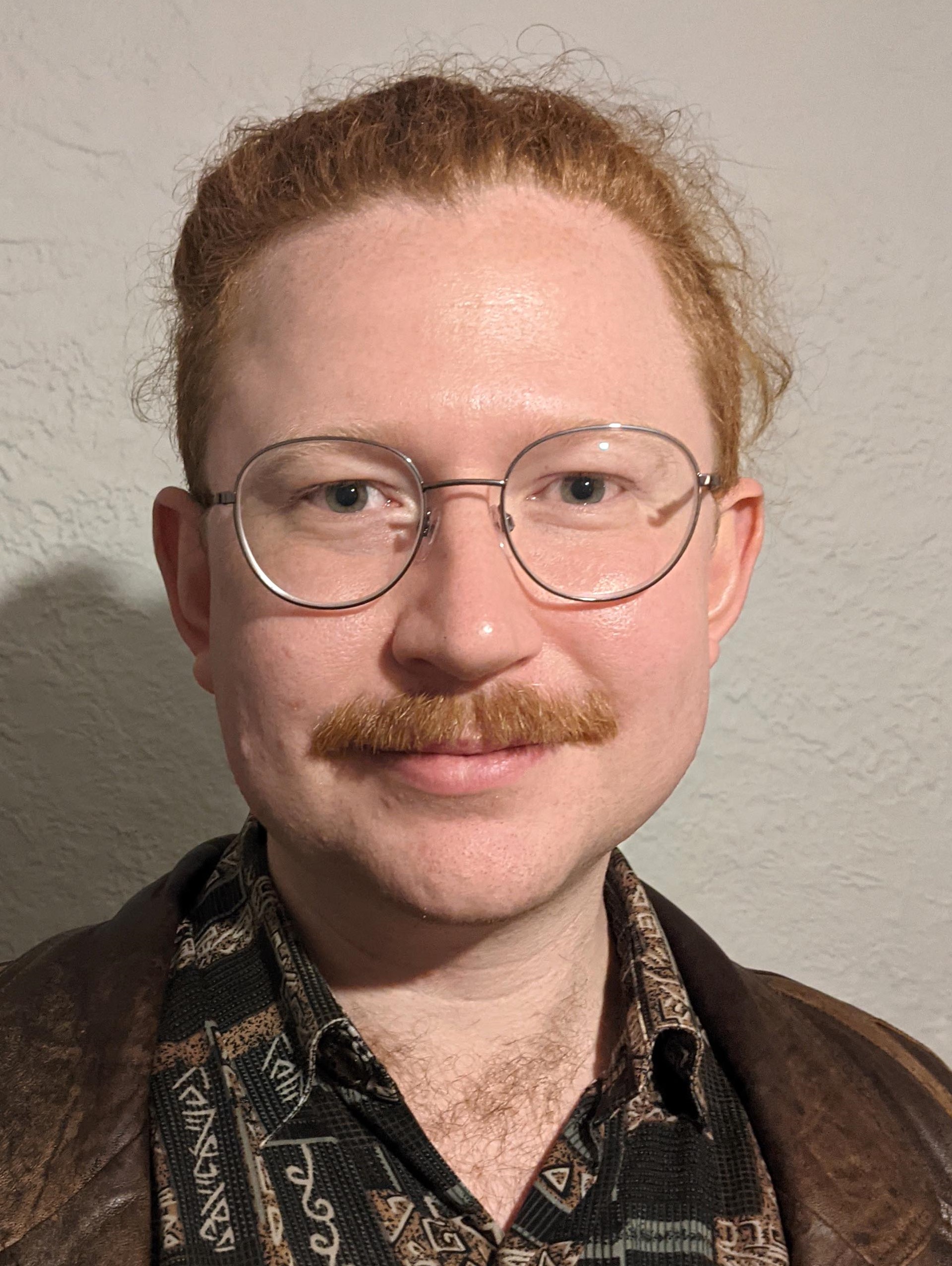}
\end{wrapfigure}
\noindent \textbf{Edric Matwiejew} is a PhD candidate at The University of Western Australia with the Quantum Information, Algorithms and Simulation (QUISA) Research Centre led by Prof. Jingbo Wang. He develops software for the high-performance simulation of quantum systems, which he applies to the design of quantum algorithms with near-term applications. In his downtime, he enjoys re-imagining scientific concepts in modular synthesizer design.

\vspace{1em}

\setlength\intextsep{0pt}
\begin{wrapfigure}{l}{2.5cm}
\raggedright
\includegraphics[width=0.3\columnwidth]{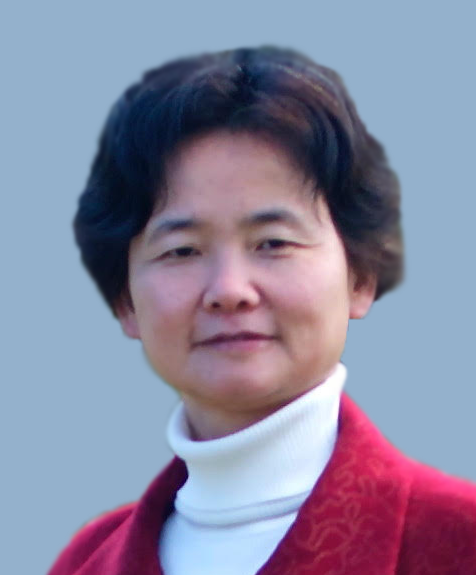}
\end{wrapfigure}
\noindent \textbf{Professor Jingbo Wang} is the Director of the QUISA Research Centre (https://quisa.tech/) hosted at The University of Western Australia, leading an active group in the area of quantum information, simulation, and algorithm development. Prof. Wang and her team pioneered quantum walk-based algorithms to solve problems of practical importance otherwise intractable, which include complex network analysis, graph theoretical studies, machine learning, and combinatorial optimisation.  She is currently also the Head of Physics Department, Deputy Head of School of Physics, Mathematics and Computing, and Chair of QST (Quantum Science and Technology) Topical Group under the Australian Institute of Physics.
\end{document}

%% file: definitions.tex
\newcommand{\bra}[1]{{\langle#1\rvert}}
\newcommand{\ket}[1]{{\lvert#1\rangle}}
\newcommand{\braket}[1]{{\langle#1\rangle}}
\newcommand{\set}[1]{{\left\{#1\right\}}} 
\newcommand{\size}[1]{{\left|#1\right|}} 

\newcommand{\quop}{{QuOp\_MPI}}
\newcommand{\python}{Python}

\newcommand{\qaoa}{QAOA}
\newcommand{\extendedQaoa}{ex-QAOA}
\newcommand{\qaoaz}{QAOAz}
\newcommand{\qwoa}{QWOA}

\newcommand{\optimalSolution}{{\Bar{s}}}
\newcommand{\solutionSpace}{{\mathcal{S}}}
\newcommand{\validSolutionSpace}{{\mathcal{S}^\prime}}
\newcommand{\validSolution}{{s^\prime}}
\newcommand{\solution}{s}
\newcommand{\solutionConstraints}{{\bm{\chi}}}
\newcommand{\solutionElements}{{\bm{\zeta}}}
\newcommand{\variationalParameters}{\bm{\theta}}

\newcommand{\finalTheta}{\boldsymbol{\theta}_f}
\newcommand{\gradient}{{\nabla_{\boldsymbol{\theta}} f}}
\newcommand{\costFunction}[1]{{C(#1)}}
\newcommand{\qualityVector}{\text{diag}(\hat{Q})}
\newcommand{\qualityOperator}{\hat{Q}}

\newcommand{\constraintFunction}[2]{{\chi_{\text{#1}}(#2)}}

\newcommand{\Uindex}{{U^{\dag}_{\#}}}

\newcommand{\ansatzFinal}[1]{{\ket{\finalTheta}_\text{#1}}}
\newcommand{\ansatz}[1]{{\ket{\variationalParameters}_\text{#1}}}
\newcommand{\initialState}[1]{{\ket{\psi_0}_\text{#1}}}
\newcommand{\U}[1]{{\hat{U}_{\text{#1}}}}
\newcommand{\finalValue}[1]{\objectiveFunction{#1}(\finalTheta)}
\newcommand{\objectiveFunction}[1]{f_\text{#1}}

\newcommand{\mpiRank}{\text{rank}}
\newcommand{\mpiAlg}{\text{COMM-}{\ansatz{}}}
\newcommand{\mpiJac}{\text{COMM-}{\nabla_{\variationalParameters}\objectiveFunction{}}}
\newcommand{\mpiSize}{\text{size}}

\newcommand{\mpiLocalElements}{\text{local}}

%% file: mpi-theta.tikz
\begin{tikzpicture}[node distance = 0.2cm, circle dotted/.style={dash pattern=on .05mm off 1.5mm, line cap=round}]

    \draw (2, 2) node[scale=4] (unitary-mpi) {$\hat{\text{U}}(\boldsymbol{\theta})$};
    
    \draw (1.2, 1.1) node[scale=1] {(MPI)};
    
    \draw (2,-0.7) node[scale=1] (unitary) {$\U{}$};
    \draw[-Latex] let \p1 = (unitary.north) in (\p1) -- (\x1, -0.05); 
    
    \draw[thick] (0.05, -0.15) -- (-0.15, -0.15) -- (-0.15, 4.15) -- (0.05, 4.15);
   
    \draw[dashed] (0,4) -- (4,4) -- (4,0) --  (0,0) -- cycle;
     
    \draw [thick] (3.8, -0.15) -- (4.15, -0.15) -- (4.15, 4.15) -- (3.8, 4.15);
   
    \draw (4.3,0) -- (4.3,0.95) -- (4.5,0.95) -- (4.5,0) -- cycle;
    
    \draw[line width = 0.7mm, circle dotted] (4.4,1.25) -- (4.4,1.65);
    
    \draw (4.3,2.05) -- (4.3,2.95) -- (4.5,2.95) -- (4.5,2.05) -- cycle;
    
    \draw (4.3,3.05) -- (4.3,3.95) -- (4.5,3.95) -- (4.5,3.05) -- cycle;
    
    \draw (4.4, -0.7) node (initial) {$\initialState{}$};
    
    \draw[-Latex] (initial.north)--(4.4,-0.05);
    
    \draw (5.1,2) node[scale=2] {\textbf{=}};
    
    \draw (5.6,0) -- (5.6,0.95) -- (5.8,0.95) -- (5.8,0) -- cycle;

    \draw[line width = 0.7mm, circle dotted] (5.7,1.25) -- (5.7,1.65);
        
    \draw (5.6,2.05) -- (5.6,2.95) -- (5.8,2.95) -- (5.8,2.05) -- cycle;
    
    \draw (5.6,3.05) -- (5.6,3.95) -- (5.8,3.95) -- (5.8,3.05) -- cycle;
    
    \draw (5.7, -0.7) node (final) {$\ansatz{}$};
    \draw[-Latex] (final.north)--(5.7,-0.05);
    
    \draw[thick,decorate, decoration = brace] (6,3.95)--(6,3.05);

    \draw (6.6,3.5) node {$\mpiLocalElements_0$};
     
    \draw[thick,decorate, decoration = brace] (6,2.95)--(6,2.05);
    
    \draw (6.6,2.5) node {$\mpiLocalElements_1$};
     
    \draw[thick,decorate, decoration = brace] (6,0.95)--(6,0.05);
    
    \draw (6.6,0.5) node {$\mpiLocalElements_2$};
    
    \draw[thick, decorate, decoration = {calligraphic brace, amplitude = 8pt}] (-0.3, 4.4) -- (6, 4.4);
   
    \draw (2.8, 5.045) node[scale=1.5] {$\mpiAlg$};
    
\end{tikzpicture}

%% file: mpi-jac.tikz
\begin{tikzpicture}[every text node part/.style={align=center}, node distance=0.7cm and 1cm, circle dotted/.style={dash pattern=on .05mm off 3.0mm, line cap=round}]
    \node[scale=1.1, draw, style = {rounded corners}, text width = 3.2cm]  (zero) {$\objectiveFunction{}(\variationalParameters)$};
    \node[scale=1.2, above = of zero, draw, style = {rounded corners}] (two) {$\left\{\frac{\partial \objectiveFunction{}(\boldsymbol{\theta})}{\partial \theta_j} \right\}$\\$2$};
    \node[scale = 1.2, left = of two, draw, style = {rounded corners}] (one) {$\left\{\frac{\partial\objectiveFunction{}(\boldsymbol{\theta})}{\partial \theta_i} \right\}$\\$1$};
    \node[scale = 1.2, right = of two, draw, style = {rounded corners}] (three) {$\left\{\frac{\partial \objectiveFunction{}(\boldsymbol{\theta})}{\partial \theta_k} \right\}$\\$m$};
    
    \draw[Latex-Latex, dashed] (one.south) -- (zero.west);
    \draw[Latex-Latex, dashed] (two.south) -- (zero.north);
    \draw[Latex-Latex, dashed] (three.south) -- (zero.east);
    
    \draw[thick, decorate, decoration = {calligraphic brace, amplitude = 8pt}] let \p1 = (one.west), \p2 = (three.east) in (\x1, \y1 + 25) -- (\x2, \y1 + 25);
   
    \draw let \p1=(two.north) in (\x1, \y1 + 28) node[scale=1.5] {$\mpiJac$};
    
    \draw[line width = 0.8mm, circle dotted] let \p1=(two.east), \p2=(three.west) in (\x1 + 3, \y1) -- (\x2, \y2);
    
\end{tikzpicture}

%% file: unitary_class_table.tex
\begin{table}[p!]
\centering
\begin{tcolorbox}[enhanced, boxsep=5pt, left = 0pt, right=0pt, top=4pt, bottom=4pt, width=\columnwidth, colframe=black!100!white!20, colback=white, coltitle = black]

\begin{tikzpicture}[node distance=0.1cm and 0.3cm, method/.style = {rectangle, text width= 0.97\columnwidth, fill=black!40!white!10}]

    \node[method] (plan) {
    \textbf{plan(self, int N, Intracomm MPI\_COMM)}: 
    \vspace{-0.1cm}
    \begin{quote}
        Determine the $\mpiAlg$ parallel partitioning scheme of $\ansatz{}$ over \verb|Intracomm MPI_COMM| for a system of size \verb|int N| and allocate memory required by external libraries (e.g. C, C++, or Fortran subroutines) called via the \verb|propagate| method.
    \end{quote}
    };
        
    \node[method] [below = of plan] (copyplan) {
    \textbf{copy\_plan(self, unitary u)}: 
    \vspace{-0.1cm}
    \begin{quote} 
    Copy a parallel partitioning scheme from the \verb|int local_i|, \verb|int local_i_offset| and \verb|array(int) partition_table| attributes of a planned \verb|unitary|. Carry out planning tasks as described above.
    \end{quote}
    };
    
    \node[method] [below = of copyplan] (propagate) {\textbf{propagate(self, float/array(float) thetas)}: 
    \vspace{-0.1cm}
    \begin{quote} 
        Compute the action of a $\U{phase}$ or $\U{mix}$ in MPI parallel over $\mpiAlg$ given input $\theta_i$ and attributes \verb|Intracomm MPI_COMM|, \verb|array(complex) initial_state| (the partitioned $\ket{\psi}$), \verb|array(complex) final_state| (the partitioned $\ansatz{}$) and matrix exponents(s) \verb|obj operator|.    
    \end{quote}
    };
    
    \node[method] [below = of propagate] (destroy) {\textbf{destroy(self)}: 
    \vspace{-0.1cm}
    \begin{quote} 
        End background processes and free memory allocated in \verb|plan| or \verb|copy_plan| that are  not managed by the \python{} garbage collector.
    \end{quote}
    };
    
\end{tikzpicture}
\end{tcolorbox}
\cprotect\caption{Unimplemented methods in the \verb|Unitary| class. A backend for parallel evaluation of a $\U{phase}$ or $\U{mix}$ unitary is incorporated into \quop{} through implementation of these methods in an \verb|unitary| subclass. Attribute types \verb|array| and \verb|Intracomm| are defined as in \cref{tab:functions}. Type \verb|complex| refers to the NumPy numerical type for double-precision complex numbers \verb|numpy.complex128|.}
\label{tab:unitary}
\end{table}

%% file: functions_table.tex
\begin{table}[pt!]
\centering
\begin{tcolorbox}[enhanced, boxsep=4pt, left = 0pt, right=0pt, top=1pt, bottom=4pt, width=\columnwidth, colframe=black!100!white!20, colback=white, coltitle = black]

\begin{tikzpicture}[node distance=0.1cm and 0.0cm, method/.style = {rectangle, text width=0.97\columnwidth, fill=black!40!white!10}]

    \node[method] (operator) {\textbf{Function:} \verb|operator|
    \\
    \textbf{Associated class:} \verb|Unitary|
    \\
    \textbf{Binds to attributes:} \verb|int system_size|, \verb|int local_i|, \verb|array(int) partition_table|, \verb|array(float) variational_parameters|, \verb|int seed|, \verb|Intracomm MPI_COMM|.  
    \\
    \textbf{Description:} Implements parallel generation of a mixing or phase-shift matrix exponent for a given \verb|unitary| state propagation method. The matrix \verb|obj operator| may be constant or parameterised by an arbitrary number of $\theta_i$ passed via the  \verb|variational_parameters| attribute. In the latter case, \verb|operator| is called with each update of $\theta_i$.};
        
    \node[method] [below = of operator] (param) {\textbf{Function:} \emph{param}
    \\
    \textbf{Associated class:} \verb|Unitary|
    \\
    \textbf{Binds to attributes:}  \verb|int system_size|, \verb|obj operator|, \verb|int n_params|, \verb|int seed|, \verb|Intracomm MPI_COMM|. 
    \\
    \textbf{Description:} Generates initial $\theta_i$ for a \verb|unitary| instance. Required positional argument \verb|n_params| specifies the size of the associated $\size{\theta_i}$. The \verb|obj operator| attribute references the matrix exponent returned by the bound \verb|operator| function.  
    };
    
    \node[method] [below = of param] (observable) {\textbf{Function:} \verb|observable|
    \\
    \textbf{Associated class:} \verb|Ansatz|
    \\
    \textbf{Binds to Attributes:} \verb|int system_size|, \verb|int local_i|, \verb|int local_i_offset|, \verb|array(int) partition_table|,\verb|Intracomm MPI_COMM|
    \\
    \textbf{Description:} Implements parallel generation of $\qualityVector$, returning a local vector partition as an \verb|array(float)| of size \verb|local_i| with a global offset of \verb|local_i_offset|. 
    };

    \node[method] [below = of observable] (state) {\textbf{Function:} \verb|state|
    \\
    \textbf{Associated class:} \verb|Ansatz|
    \\
    \textbf{Binds to attributes:} \verb|int system_size|, \verb|int local_i|, \verb|int local_i_offset|, \verb|array(int) partition_table|, \verb|Intracomm MPI_COMM|
    \\
    \textbf{Description:} Implements parallel generation of $\initialState{}$, returning a local vector partition as described above.};

\end{tikzpicture}
\end{tcolorbox}
\cprotect\caption{\quop{} function types. Passed to the \verb|Ansatz| class by the corresponding `set' method (see \cref{tab:ansatz}), positional arguments are mapped to attributes of the either the \verb|Unitary| and \verb|Ansatz| classes. An arbitrary number of keyword arguments are permitted. Attributes \verb|local_i|, \verb|local_i_offset| and \verb|partition_table| are defined in accordance with \cref{eq:local_i,eq:offset} with \verb|partition_table| containing the global start and end partition indexes of the distributed state vector. Integer \verb|seed| is provided for reproducible pseudo-random number generation. Type \verb|array| refers to a 1-dimensional NumPy ndarray and type \verb|Intracomm| refers to an MPI4Py MPI intra-communicator. }
\label{tab:functions}
\vspace{1em}
\end{table}

%% file: ansatz_class_table.tex
\begin{table*}[t!]
    \centering
    
        \begin{tcolorbox}[enhanced, boxsep=0pt, left = 5pt, right=0pt, top=5pt, bottom=5pt, width=0.955\textwidth, colframe=black!100!white!20, colback=white, coltitle = black]
        
            \begin{tikzpicture}[node distance=1.1cm and 0.15cm, set/.style = {rectangle, fill=black!40!white!10, minimum width=7.5cm, text width= 8cm}, gen/.style = {rectangle, fill=black!40!white!10, minimum width=7.5cm,  text width= 8.5cm}, sim/.style = {rectangle, fill=black!40!white!10, minimum width=7.5cm, text width= 8.5cm}, get/.style = {rectangle, fill=black!40!white!10, minimum width=7.5cm, text width= 8.5cm}, save/.style = {rectangle, fill=black!40!white!10, minimum width=7.5cm, text width= 8.5cm},  print/.style = {rectangle, fill=black!40!white!10, minimum width=7.5cm, text width= 8.5cm}]
            
                \node[set] [anchor = west] (setunitaries) {
                \textbf{set\_unitaries} [required]:
                \vspace{-0.1cm}
                \begin{quote}
                	Define $\U{}$ via a list of \verb|unitary| instances.
                \end{quote}
                };

                \node[set] [below = of setunitaries.west, anchor = west] (setobservables) {\textbf{set\_observables} [required]:
                \vspace{-0.1cm}
                \begin{quote}
                    Define $\qualityVector$ via a \verb|quality| function.
                \end{quote}
                };
                
                \node[set] [below = of setobservables.west, anchor = west] (setinitialstate) {\textbf{set\_initial\_state} [default $\initialState{}$ = $\ket{+}$]:
                \vspace{-0.1cm}
                \begin{quote}
                    Define $\initialState{}$ via a \verb|state| function.
                \end{quote}
                };
                
                \node[set] [below = of setinitialstate.west, anchor = west] (setdepth) {\textbf{set\_depth} [default $D = 1$]:
                \vspace{-0.1cm}
                \begin{quote}
                    Define $D$.
                \end{quote}
                };
                
                \node[set] [below = of setdepth.west, anchor = west] (setoptimiser) {\textbf{set\_optimiser} [default Scipy BFGS, $\text{tol} = 1^{-5}$]:
                \vspace{-0.1cm}
                \begin{quote}
                    Specify the classical optimiser.
                \end{quote}
                };
                
                \node[set] [below = of setoptimiser.west, anchor = west] (setseed) {\textbf{set\_seed} [default $\text{seed} = 0$]:
                \vspace{-0.1cm}
                \begin{quote}
                    Specify a random seed for \quop{} functions.
                \end{quote}
                };
                
                \node[set] [below = of setseed.west, anchor = west] (setparallel) {\textbf{set\_parallel} [default $\mpiAlg$ only]:
                \vspace{-0.1cm}
                \begin{quote} 
                    Specify the parallelisation scheme.
                \end{quote}
                };
                
                \node[set] [below = of setparallel.west, anchor = west] (setlog) {\textbf{set\_log} [optional]:
                \vspace{-0.1cm}
                \begin{quote} 
                    Specify simulation data-logging.
                \end{quote}
                };
                
                \node[set] [below = of setlog.west, anchor = west] (setobservablemap) {\textbf{(un)set\_observable\_map} [optional]:
                \vspace{-0.1cm}
                \begin{quote} 
                Define $g$ in $q = g(\qualityVector)$, where $g: \mathbb{R^N} \rightarrow \mathbb{R}^N$.
                \end{quote}
                };
                
                \node[set] [below = of setobservablemap.west, anchor = west] (setobjectivemap) {\textbf{(un)set\_objective\_map} [optional]:
                \vspace{-0.1cm}
                \begin{quote} 
                Define $h$ in $f(\variationalParameters) = h(\bra{\variationalParameters}\hat{Q} \ket{{\variationalParameters}})$, where $h: \mathbb{R} \rightarrow \mathbb{R}$.
                \end{quote}
                };
                
                \node[gen] [right = of setunitaries.east, anchor = west] (genparams) {\textbf{gen\_initial\_params}:
                \vspace{-0.1cm}
                \begin{quote} 
                    Generate and return $\variationalParameters_0$.
                \end{quote}
                };
                
                \node[sim] [below = of genparams.west, anchor = west] (evolvestate) {\textbf{evolve\_state}:
                \vspace{-0.1cm}
                \begin{quote} 
                    Compute $\ansatzFinal{}$ for input $\variationalParameters_0$.
                \end{quote}
                };
                
                \node[sim] [below = of evolvestate.west, anchor = west] (execute) {
                \textbf{execute}:
                \vspace{-0.1cm}
                \begin{quote} 
                    Minimise $f(\variationalParameters)$
                \end{quote}
                };
                
                \node[sim] [below = of execute.west, anchor = west] (benchmark) {
                \textbf{benchmark}:
                \vspace{-0.1cm}
                \begin{quote} 
                    Minimise $f(\variationalParameters)$ over $D =  (d_\text{min},...,d_\text{max})$.
                \end{quote}
                };
                
                \node[get] [below = of benchmark.west, anchor = west] (finalstate) {
                \textbf{get\_final\_state}:
                \vspace{-0.1cm}
                \begin{quote} 
                    Return $\ansatzFinal{}$ at $\mpiRank = 0$ of the global MPI communicator.
                \end{quote}
                };
                
                \node[get] [below = of finalstate.west, anchor = west] (probabilities) {\textbf{get\_probabilities}:
                \vspace{-0.1cm}
                \begin{quote} 
                    Return $\braket{s_i|\variationalParameters_f}$ at $\mpiRank = 0$ of the global MPI communicator.
                \end{quote}
                };
                
                \node[get] [below = of probabilities.west, anchor = west] (expectationvalue) {\textbf{get\_expectation\_value}:
                \vspace{-0.1cm}
                \begin{quote} 
                    Return $\finalValue{}$ at $\mpiRank = 0$ of the global MPI communicator.
                \end{quote}
                };
                
                \node[save] [below = of expectationvalue.west, anchor = west] (save) {\textbf{save}:
                \vspace{-0.1cm}
                \begin{quote} 
                    Write $\ansatz{}$, $\qualityVector$, $\variationalParameters$ and the optimisation result to disk.
                \end{quote}
                };
                
            \node[print] [below = of save.west, anchor = west] (print) {
                \textbf{print\_optimiser\_result}:
                \vspace{-0.1cm}
                \begin{quote} 
                    Print the result summary of the classical optimiser.
                \end{quote}
                };
                
            \end{tikzpicture}
        \end{tcolorbox}
        \cprotect\caption{An overview of the public \verb|Ansatz| class methods. Methods tagged as `required' must be called before initialisation of QVA state propagation via the \verb|evolve_state|, \verb|execute| or \verb|benchmark| methods.}
        \label{tab:ansatz}
\end{table*}